\documentclass[amsfonts,amsmath,amssymb,twocolumn,pre]{revtex4}
\usepackage{graphicx,subfigure}
\usepackage{times}

\begin{document}

\title{Concentration and mass dependence of transport coefficients and correlation functions in binary mixtures with high mass-asymmetry}
\author{W. Fenz$^{1}$, I. M. Mryglod$^{1,2}$, O. Prytula$^{1,2}$, R. Folk$^{1}$}
\affiliation{$^{1}$Institute for Theoretical Physics, Linz University,
A-4040 Linz, Austria\\
$^{2}$Institute for Condensed Matter Physics, 1 Svientsitskii Street, UA-79011 Lviv, Ukraine}
\date{\today}
\begin{abstract}
Correlation functions and transport coefficients of self-diffusion and shear viscosity of a binary
Lennard-Jones mixture with components differing only in their particle mass
are studied up to high values of the mass ratio $\mu$, including the limiting case $\mu=\infty$, for different mole fractions $x$. Within a large range of $x$ and $\mu$ the product of the diffusion coefficient of the heavy species $D_{2}$ and the total shear viscosity of the mixture $\eta_{m}$ is found to remain constant, obeying a generalized Stokes-Einstein relation.
At high liquid density, large mass ratios lead to a pronounced cage effect that is observable in the mean square displacement, the velocity autocorrelation function and the van Hove correlation function.

\vspace{6pt}
\end{abstract}
\maketitle

\section{Introduction}

The dynamic properties of binary fluid systems where one particle species differs from the other only in size, mass or both of these parameters have been the subject of a large number of studies during the last years \cite{espanol,nuevo,nuevo2,ould,ould2,ould3,Schmidt,soko,McPhie,McPhie2,lee,ali}. The increasing interest is, on the one hand, due to the fact that such systems serve as simple models for colloids and micellar solutions, which are of prime importance in many scientific areas such as biology or biochemistry, on the other hand it is sparked by the rapidly growing capabilities of modern computer hardware which allows us to investigate parameter ranges and system sizes that were not accessible before.

In general, there are two important limiting cases that can be focused on. The first and best-investigated so far is the so-called \emph{'tracer'} or \emph{Brownian} limit of one single heavy and/or large molecule suspended in a solvent of light particles (infinite dilution). This is especially interesting because for the case of a macroscopically sized and in comparison to the solvent infinitely heavy tracer particle, there is a simple relation between the tracer diffusivity $D$ and the viscosity of the solvent $\eta$. This \emph{Stokes-Einstein} (SE) relation states that
    \[D=\frac{k_BT}{C\eta},
\]
where $k_B$ denotes Boltzmann's constant, $T$ the temperature, and $C$ is a numeric constant depending on geometric boundary conditions. Several studies have been devoted to the Brownian limit \cite{espanol,nuevo,ould,ould3,soko,lee}, some of them especially to the question, for which range of tracer mass and size this purely hydrodynamic relation also holds on the microscopic level \cite{ould2,Schmidt}. Depending on whether the mass is changed at constant size ratio or not, the SE relation was found to hold for mass ratios larger than 10 \cite{ould2} and larger than 100 \cite{Schmidt}, respectively. Above these values, the tracer diffusion was considered mass-independent. (In this context, it is also interesting to note that even for a pure simple fluid the SE relation was found to hold in a large part of the phase diagram \cite{cap}.)

Considerably less studies exist on the approach to the Brownian limit (small but finite concentration of heavy particles, \cite{nuevo2,McPhie,McPhie2}) and mixtures with larger mole fractions of the heavy component \cite{ali}.

The second limiting case corresponds to the scenario when the mass of the heavy species goes to infinity. For the case of a single Brownian particle, this situation was covered in \cite{espanol}, where it is explained why it makes sense to attribute a finite friction coefficient to an immobile particle.
For a finite concentration of heavy particles, on the other hand, the infinite-mass limit effectively transforms the system into a one-component fluid in a random porous matrix of fixed obstacles that takes up a finite fraction of volume. As a consequence, there exists a percolation transition at large density and concentration of the heavy component which ultimately prevents the light particles from diffusing through the system.

But also in the case of finite mass ratio and concentration, the heavy particles influence the dynamics of the light ones. Since different masses induce different time scales, the heavy molecules act as a cage for the light ones, stalling their diffusion until they themselves have finally moved significantly from their starting position. This leads to so-called hopping processes, characterized by particles moving from one cage to the next.
These complicated dynamic processes, which also occur near the glass transition in supercooled liquids \cite{kob,wahn}, have made it difficult to tackle such systems by theory, which is why many of the existing studies rely mostly on computer simulations. The most successful theory at present is the mode-coupling theory (MCT), which in its general form incorporates a mathematical description of hopping processes, but also in its idealized form yields already rather good results for glassy systems \cite{kob,kob2,kob3,gallo} as well as highly mass- or size-asymmetric mixtures \cite{ali}. Recently, the mean-field theory of Tokuyama has also been applied successfully to diverse glass-forming systems \cite{toku1,toku2,toku4,toku5,toku6}.

Now, the goal of our work is to study the approach to the infinite-mass limit in an asymmetric binary mixture at finite concentrations of the heavy component, using molecular dynamics calculations on as simple a model as possible. In particular, we have chosen a truncated and shifted Lennard-Jones interaction potential with a cut-off radius of $r_c=2.5$, where both species have equal interaction strength and particle size. We perform simulations at two state points away from the critical region, one with moderate and one with high liquid density (the phase diagram for this system can be found in \cite{shi}). The only two remaining system parameters, the mass ratio $\mu$ and mole fraction $x$, do not influence the static properties of the system at all. What we are interested in are the transport coefficients of self-diffusion $D_1$ and $D_2$ and the shear viscosity $\eta_m$ of the mixture, especially the dynamics for high $\mu$ including the case $\mu=\infty$. In order to avoid dealing with the percolation th
reshold, we focus on small concentrations $x\leq0.2$. Although there are no critical fluctuations, due to the periodicity of the simulation box the diffusivity is afflicted with a finite-size effect that has to be accounted for \cite{fushiki}. We also study a possible relation between $D_2$ and $\eta_m$ (generalized SE relation) and the influence of high $\mu$ on the dynamics of the light particles (cage effect).

The paper is organized as follows: In the next section, we discuss in more detail the model system and the simulation methods we apply. After that, we give a brief overview on the basic formulas for the dynamic quantities we are going to look at, and subsequently we present the results we have obtained. Finally, we end with a conclusion and a short outlook.

\section{Model and simulation details}

We consider a two-component mixture consisting of $N=N_{1}+N_{2}$ particles in
a cubic volume $V$ with periodic boundary conditions. A pair of particles $i,$
$j$ of species $\alpha$ and $\beta$, respectively, separated by a distance
$r=\left|  \mathbf{r}_{i}-\mathbf{r}_{j}\right|  ,$ interacts via a truncated
and shifted Lennard-Jones potential $\phi\left(  r\right)$, given by%
\begin{equation}
\phi\left(  r\right)  =\left\{
\begin{array}
[c]{cc}%
\phi_{LJ}\left(  r\right)  -\phi_{LJ}\left(  r_{c}\right)  , & \,\,r<r_{c}\\
0, & \,\,r>r_{c}%
\end{array}
\right.  ,
\end{equation}
and%
\begin{equation}
\phi_{LJ}\left(  r\right)  =4\varepsilon_{\alpha\beta}\left[  \left(
\frac{\sigma_{\alpha\beta}}{r}\right)  ^{12}-\left(  \frac{\sigma_{\alpha
\beta}}{r}\right)  ^{6}\right]  ,
\end{equation}
where $r_{c}=2.5,$ and the two species have the same interaction strengths
$\varepsilon_{11}=\varepsilon_{22}=\varepsilon_{12}\equiv\varepsilon$ and
particle sizes $\sigma_{11}=\sigma_{22}=\sigma_{12}\equiv\sigma,$ but
different masses $m_{2}> m_{1}.$ From the point of view of statics, such a
system is identical to a one-component Lennard-Jones fluid, the dynamic
properties, however, will of course depend on the mass ratio $\mu=m_{2}/m_{1}$
and the concentration specified by the mole fraction $x=N_{2}/N$. In the
extreme limit of $\mu\rightarrow\infty$ we are effectively dealing with a
system of $N_{1}$ particles moving in a disordered matrix of $N_{2}$ fixed
obstacles of the same size.

We performed MD simulations in the $NVT$-ensemble using a Nos\'{e}-Hoover
thermostat \cite{nose,nose2,hoover,hoover2} in the formulation of Martyna et
al. \cite{mart} and a velocity Verlet integration scheme. A multiple time step
algorithm \cite{mts} was applied in order to deal with the different time scales due to the high mass asymmetry of the two mixture components.
Simulations usually lasted at least $2\times10^6$ time steps, where one time step of the light species was chosen as $\delta t=0.005 \sqrt{m_{1}\sigma^{2}/\varepsilon}$, with equilibration times of $2\times10^5$ time steps.

Simulations for infinite mass ratio $\mu=\infty$ were performed by fixing the positions of
the heavy particles at random configurations obtained from short simulation
runs under identical thermodynamic conditions but with equal masses for both
species. Usually, the reported simulation results were averaged over 10
different configurations, which turned out to be a large enough number, since
we found the dependence on the exact configurations to be quite small. Also in
the case of finite mass ratios each result is obtained as an average of several simulation runs
in order to improve the statistics and calculate standard deviations of
the dynamic quantities.

The question how the limit $\mu\rightarrow\infty$ is to be interpreted is a
delicate one \cite{espanol}. Since the heavy particles have infinite mass and
zero velocity, their momentum $\mathbf{P}_{2}=\sum_{i=1}^{N_{2}}%
m_{2}\mathbf{v}_{i}$ is in principle undefined. However, the total momentum of
the light particles $\mathbf{P}_{1}=\sum_{i=1}^{N_{1}}m_{1}\mathbf{v}_{i}$ is known
at any time, and therefore by requiring the total momentum $\mathbf{P}$ of the
system consisting of light and heavy particles to be fixed and (by convention)
equal to zero - just the same as in the simulations with finite $\mu$ - we can
assign them the finite momentum $\mathbf{P}_{2}\left(  t\right)
=-\mathbf{P}_{1}\left(  t\right)  .$ In this way, we define the system at any
time step $t$ as the limiting case of a system where $m_{2}$ goes to infinity
and $\mathbf{v}_{i}$ goes to zero for all $i\in\{1,\ldots,N_{2}\},$ while
$\mathbf{P}_{2}$ is held constant at the value $-\mathbf{P}_{1}\left(
t\right)  .$ The same procedure was applied in \cite{espanol} to calculate the
momentum autocorrelation function of a single Brownian particle with infinite mass.

In the following, all quantities will be given in reduced units: energy is
measured in $\varepsilon,$ length in $\sigma,$ mass in $m_{1}$ and time in
$\sqrt{m_{1}\sigma^{2}/\varepsilon}.$ Temperatures are given in $\varepsilon
/k_{B},$ densities in $1/\sigma^{3}.$

\section{Theoretical background}
\subsection{Mean square displacement and diffusion coefficients}

\begin{figure}[tb]
\includegraphics[width=.95\columnwidth]{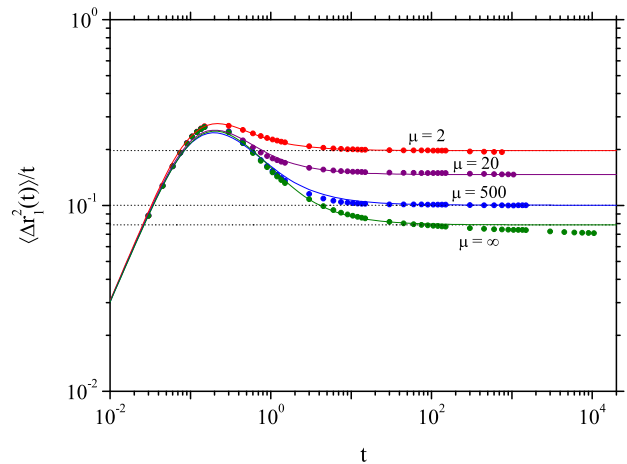} 
\caption{ (Color online) Points: MD data of the MSD of the light species divided by time for $x=0.2$
and mass ratios $\mu=2$ (red), $\mu=20$ (purple), $\mu=500$ (blue) and $\mu=\infty$ (green). Solid lines:
Mean field results, fitted for the mean free path length $l$, which is 0.15, 0.15, 0.155 and 0.164,
respectively. Dashed lines: diffusion coefficients obtained from the Green-Kubo relation, multiplied by 6.
The density is $\rho=0.9$ and temperature $T=1$.
\label{fig_MSD}}
\end{figure}

The mean square displacement (MSD) of a particle of species $\alpha$ is defined as%
\begin{equation}
\left\langle \Delta r_{\alpha}^{2}\left(  t\right)  \right\rangle =\frac
{1}{N_{\alpha}}\sum_{i=1}^{N_{\alpha}}\left\langle \left[  \mathbf{r}%
_{i}\left(  t\right)  -\mathbf{r}_{i}\left(  0\right)  \right]  ^{2}%
\right\rangle,
\end{equation}
where $\mathbf{r}_{i}\left(t\right)$ is the three-dimensional trajectory of particle $i$, and $\left\langle\cdot\right\rangle$ denotes a canonical average.
In the case of normal diffusion according to Fick's law,
the MSD obeys the so-called Einstein-Helfand relation \cite{helfand}, which states that
the self-diffusion coefficient of light or heavy particles is given by its
slope at large times according to
\begin{equation}
D_{\alpha}=\lim_{t\rightarrow\infty}\frac{\left\langle \Delta
r_{\alpha}^{2}\left(  t\right)  \right\rangle }{6t}. \label{DEH}
\end{equation}
Another way to obtain the diffusion coefficients of the two components is via their velocity autocorrelation functions (VACFs) $\psi_{\alpha}(t)$, defined as
\begin{equation}
\psi_{\alpha}\left(  t\right)  =\frac{1}{3N_{\alpha}}\sum_{i=1}^{N_{\alpha}%
}\left\langle \mathbf{v}_{i}\left(  t\right)  \mathbf{v}_{i}\left(  0\right)
\right\rangle.
\end{equation}
These allow us to calculate the self-diffusion through the Green-Kubo relation \cite{HMD}
\begin{equation}
D_{\alpha}=\int_{0}^{\infty}\psi_{\alpha}\left(  t\right)  dt. \label{DGK}
\end{equation}

\begin{figure}[tb]
\subfigure[]{\includegraphics[width=.9\columnwidth]{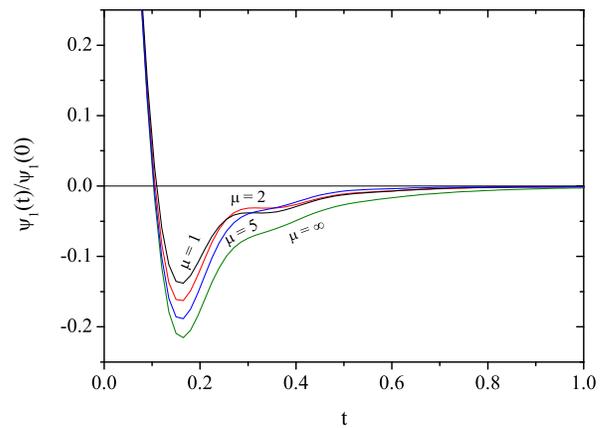}} 
\hfill\subfigure[]{\includegraphics[width=.9\columnwidth]{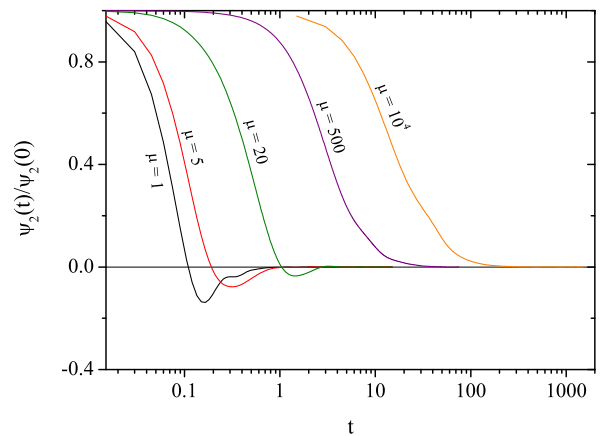}} 
\caption{ (Color online) Normalized velocity auto-correlation function $\psi_{1}(t)$ of the light (a)
and $\psi_{2}(t)$ of the heavy particles (b) in a system with $T=1$, $\rho=0.9$, $x=0.2$ and different mass ratios $\mu$.
\label{fig_vacf}}%
\end{figure}

\subsection{Mean Field Theory}

Tokuyama has established a mean field theory \cite{toku1} for the MSD near the glass transition in colloidal suspensions and molecular
systems. In three dimensions, the MSD denoted by $M\left(
t\right)  $ is described by the nonlinear differential equation%
\begin{equation}
\frac{d}{dt}M\left(  t\right)  =6D+6\left[  v_{0}^{2}t-D\right]
e^{-M\left(  t\right)  /l^{2}},
\end{equation}
with the formal solution%
\begin{align}
M\left(  t\right)   &  =6Dt+l^{2}\ln\left\{  e^{-6Dt/l^{2}}\right. \nonumber\\
&  \left.  +\frac{l^{2}}{6D^{2}}\left[  1-\left(  1+6Dt/l^{2}\right)
e^{-6Dt/l^{2}}\right]  \right\}  . \label{M2}%
\end{align}
Here, $D$ is the self-diffusion coefficient, $l$ is the mean free path, and
$v_{0}$ denotes the average velocity of an atom. We have applied this theory
to the MSD data for the light particles from our MD
simulations, taking the value of the diffusion coefficient from the Green-Kubo
results, and fitting the value of $l$ in Eq. (\ref{M2}) to the MD data.

\subsection{Van Hove correlation function}

\begin{figure}[tb]
\includegraphics[width=.9\columnwidth]{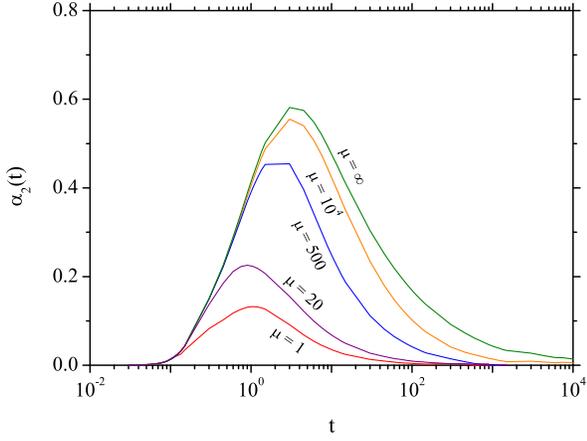}
\caption{ (Color online) Non-Gaussian parameter $\alpha_2(t)$ of the light species for a concentration of $x=0.2$ and different mass ratios.
The density is $\rho=0.9$ and temperature $T=1$.
\label{fig_alpha2}}%
\end{figure}

\begin{figure}[tb]
\subfigure[]{\includegraphics[width=.49\columnwidth]{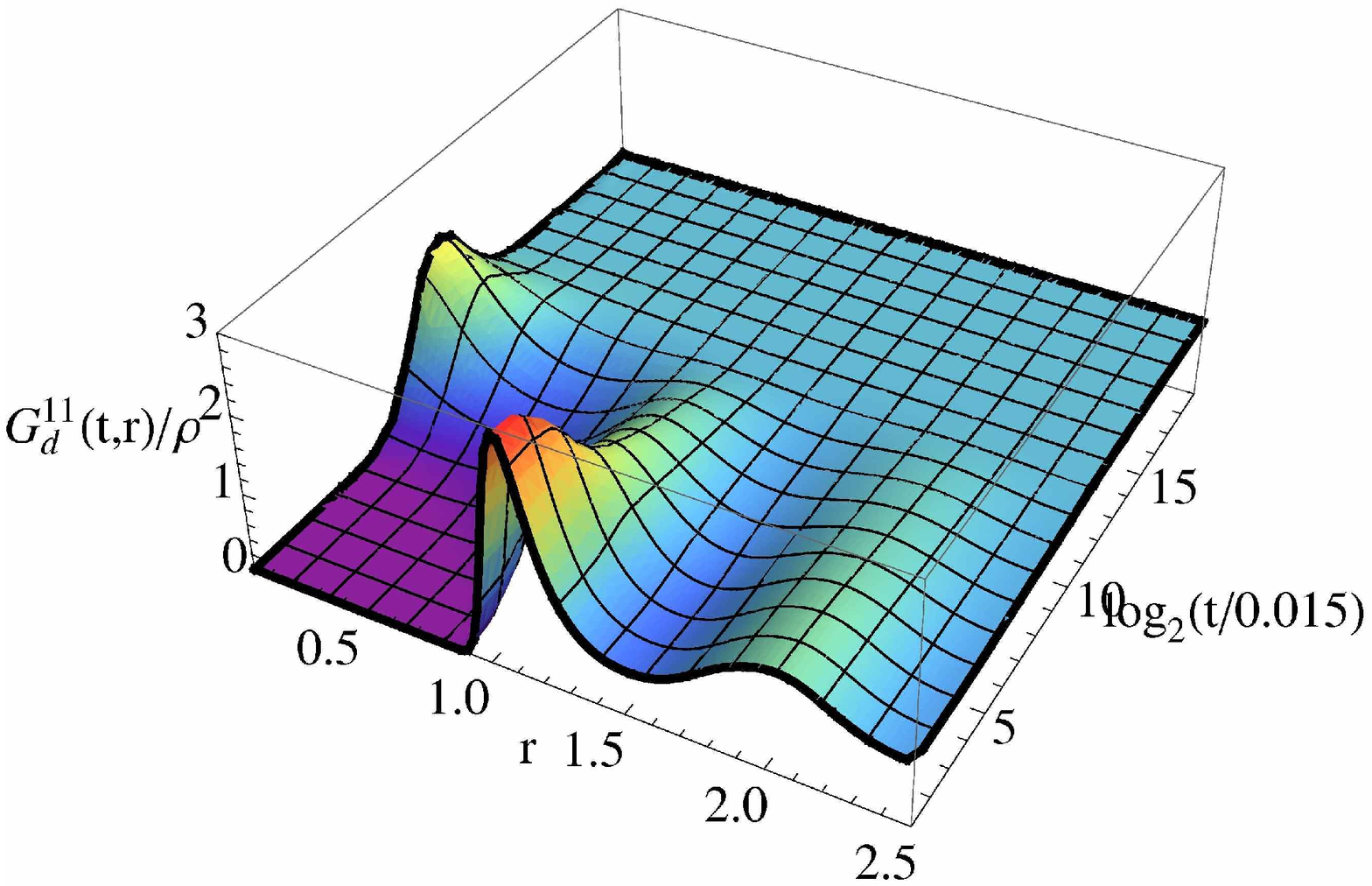}} 
\hfill
\subfigure[]{\includegraphics[width=.49\columnwidth]{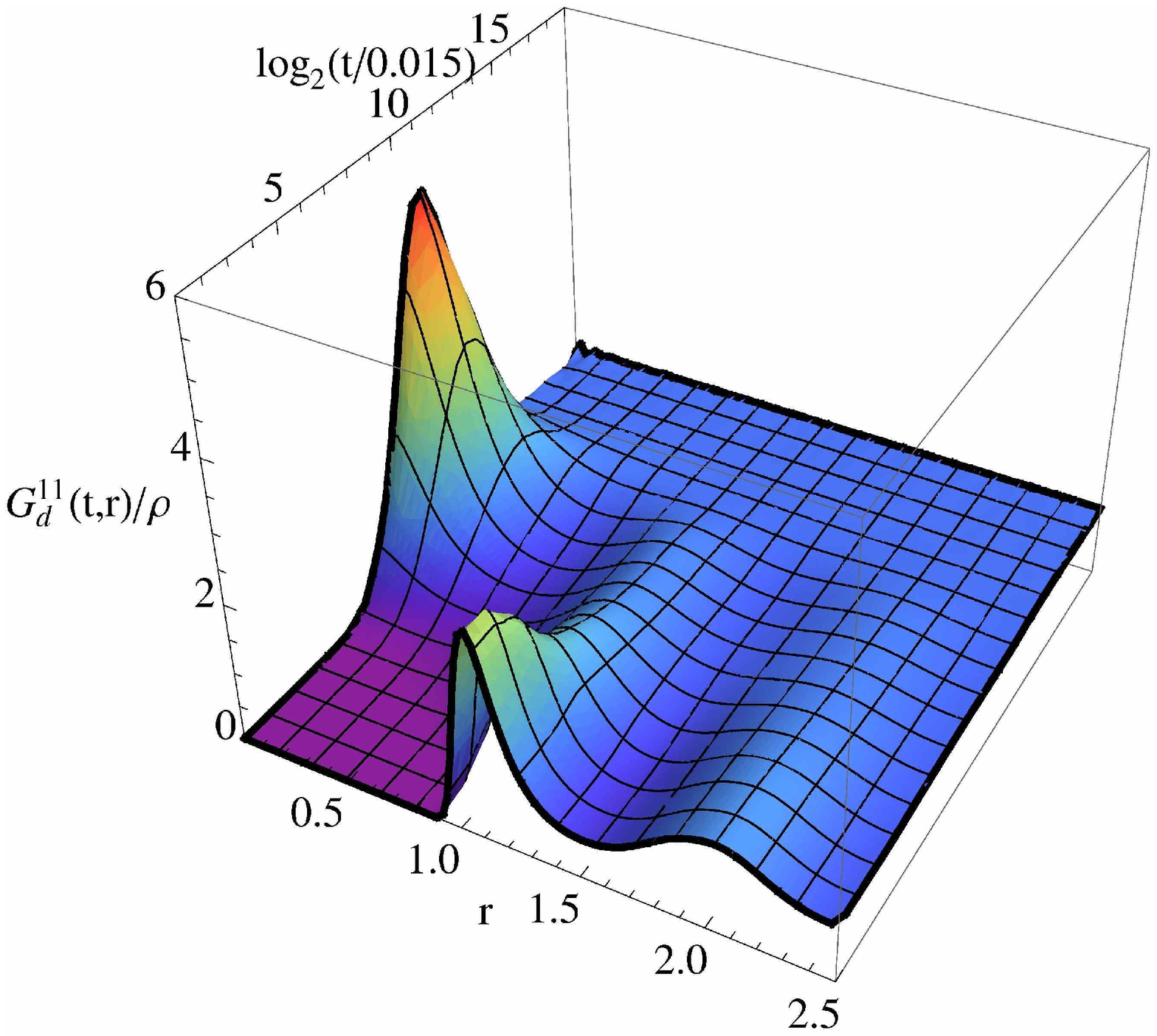}} 

\subfigure[]{\includegraphics[width=.5\columnwidth]{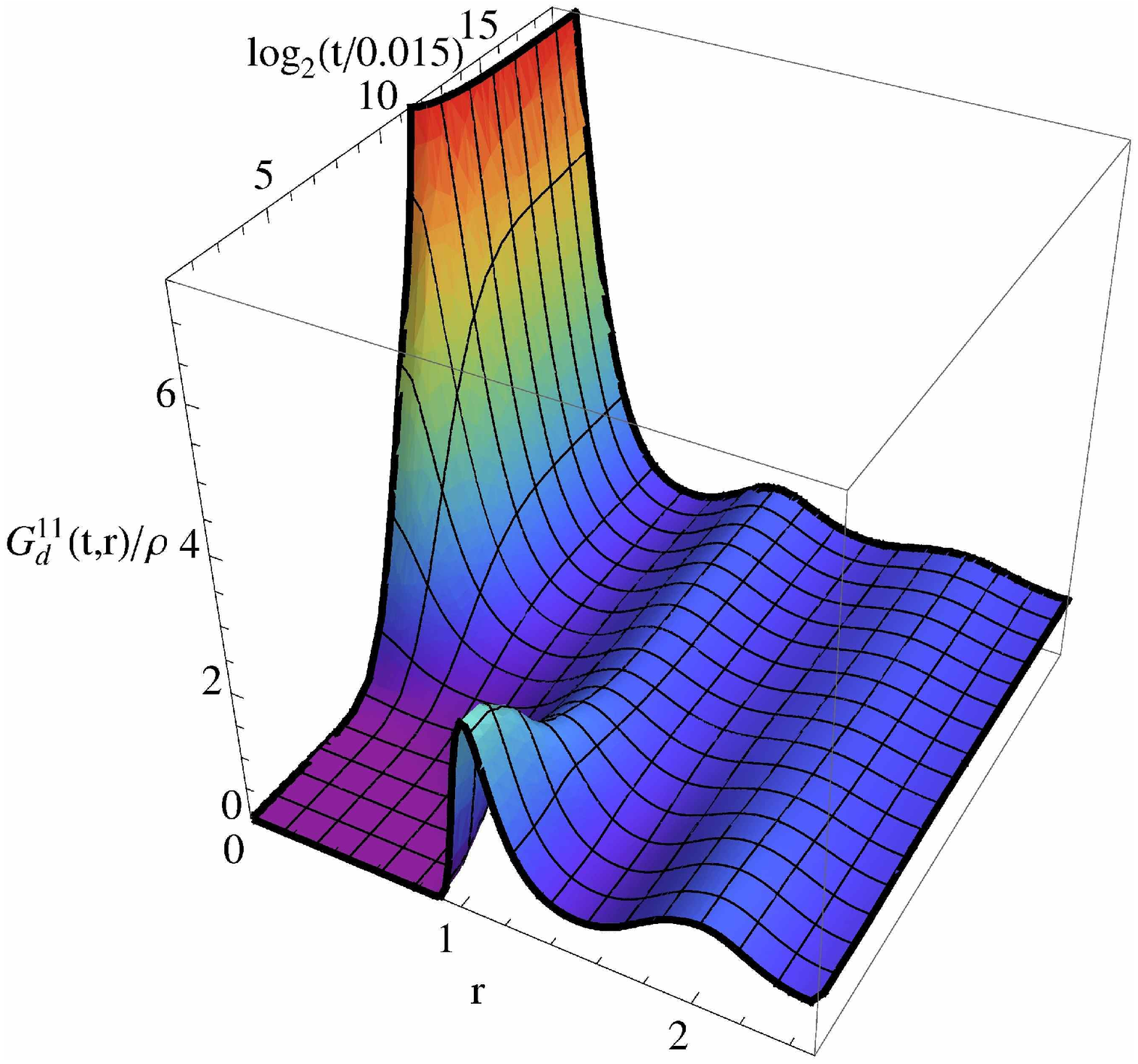}}  
\caption{ (Color online) Normalized distinct part of the van Hove function $G_d^{11}(r,t)/\rho$. The concentration of the heavy particles is $x=0.2$, and the mass ratio is $\mu=10^4$ (a), $\mu=3\times10^5$ (b) and $\mu=\infty$ (c). The density is $\rho=0.9$ and temperature $T=1$ in all cases.
\label{fig_Gd11}}%
\end{figure}

Space-time correlations between two particles in a pure fluid system are described
by the van Hove correlation function (VHCF) $G\left(  \mathbf{r},t\right)  ,$ which is defined as
\cite{vanhove}%
\begin{equation}
G\left(  \mathbf{r},t\right)  =\frac{1}{N}\left\langle \sum_{i,j=1}^{N}%
\delta\left[  \mathbf{r}+\mathbf{r}_{i}\left(  0\right)  -\mathbf{r}%
_{j}\left(  t\right)  \right]  \right\rangle .
\end{equation}
For an isotropic system, $G\left(  \mathbf{r},t\right)  d^{3}r=4\pi
r^{2}G\left(  r,t\right)  dr$ gives the probability to find a particle at time
$t$ a distance $r$ from the origin, provided that at time $t=0$ a particle was
located at the origin. The VHCF can be separated into a self- and
a distinct part,%
\begin{equation}
G\left(  r,t\right)  =G_{s}\left(  r,t\right)  +G_{d}\left(  r,t\right)  ,
\end{equation}
where $G_{s}$ includes the terms with $i=j,$ and $G_{d}$ those with $i\neq j.$
The self-part, on the one hand, is the time-dependent conditional probability
density that a particle moves a distance $r=\left|  \mathbf{r}\left(
0\right)  -\mathbf{r}\left(  t\right)  \right|  $ during time $t.$ At $t=0,$
$G_{s}\left(  r,0\right)  =\delta\left(  \mathbf{r}\right)  $, whereas
$\lim_{r\rightarrow\infty}G_{s}\left(  r,t\right)  =\lim_{t\rightarrow\infty
}G_{s}\left(  r,t\right)  =1/V\approx0.$ It is normalized to unity, $4\pi\int
r^{2}G_{s}\left(  r,t\right)  dr=1,$ and connected to the MSD via the relation%
\begin{equation}
\left\langle \Delta r^{2}\left(  t\right)  \right\rangle =\int r^{2}%
G_{s}\left(  r,t\right)  d^{3}r. \label{MSDVHCF}
\end{equation}
For large enough $r$ and $t$, the self-part approaches a Gaussian distribution whose width grows with $\sqrt{Dt}$, where $D$ is the diffusion coefficient of the system,
\begin{equation}
G_{s}\left(  r,t\right)  \overset{t,r\rightarrow\infty}{\rightarrow}\frac
{1}{\left(  4\pi Dt\right)  ^{3/2}}\exp\left(  -\frac{r^{2}}{4Dt}\right). \label{GaussVHCF}
\end{equation}

The distinct part, on the other hand, represents the conditional probability
density of finding a particle at time $t$ a distance $r$ apart from the
location of another particle at time $t=0.$ At $t=0,$ $G_{d}\left(
r,0\right)  =\rho g\left(  r\right)  $, and $\lim_{r\rightarrow\infty}%
G_{d}\left(  r,t\right)  =\lim_{t\rightarrow\infty}G_{d}\left(  r,t\right)
=\rho.$ The normalization is $4\pi\int r^{2}G_{d}\left(  r,t\right)  dr=N-1.$

In a binary mixture one has to differentiate between the different species
$\alpha$, and the corresponding VHCFs are defined as%
\[
G_{s}^{\alpha}\left(  \mathbf{r},t\right)  =\frac{1}{N_{\alpha}}\left\langle
\sum_{i=1}^{N_{\alpha}}\delta\left[  \mathbf{r}+\mathbf{r}_{i}\left(
0\right)  -\mathbf{r}_{i}\left(  t\right)  \right]  \right\rangle ,
\]
and%
\begin{equation}
G_{d}^{\alpha\alpha}\left(  \mathbf{r},t\right)  =\frac{N_{1}+N_{2}}%
{N_{\alpha}\left(  N_{\alpha}-1\right)  }\left\langle \sum_{\substack{i=1\\j\neq i}}^{N_{\alpha}%
}\delta\left[  \mathbf{r}+\mathbf{r}_{i}\left(
0\right)  -\mathbf{r}_{j}\left(  t\right)  \right]  \right\rangle ,
\end{equation}%
\begin{equation}
G_{d}^{12}\left(  \mathbf{r},t\right)  =\frac{N_{1}+N_{2}}{N_{1}N_{2}%
}\left\langle \sum_{i=1}^{N_{1}}\sum_{j=1}^{N_{2}}\delta\left[  \mathbf{r}%
+\mathbf{r}_{i}\left(  0\right)  -\mathbf{r}_{j}\left(  t\right)  \right]
\right\rangle .
\end{equation}

\begin{figure}[tb]
\subfigure[]{\includegraphics[width=.49\columnwidth]{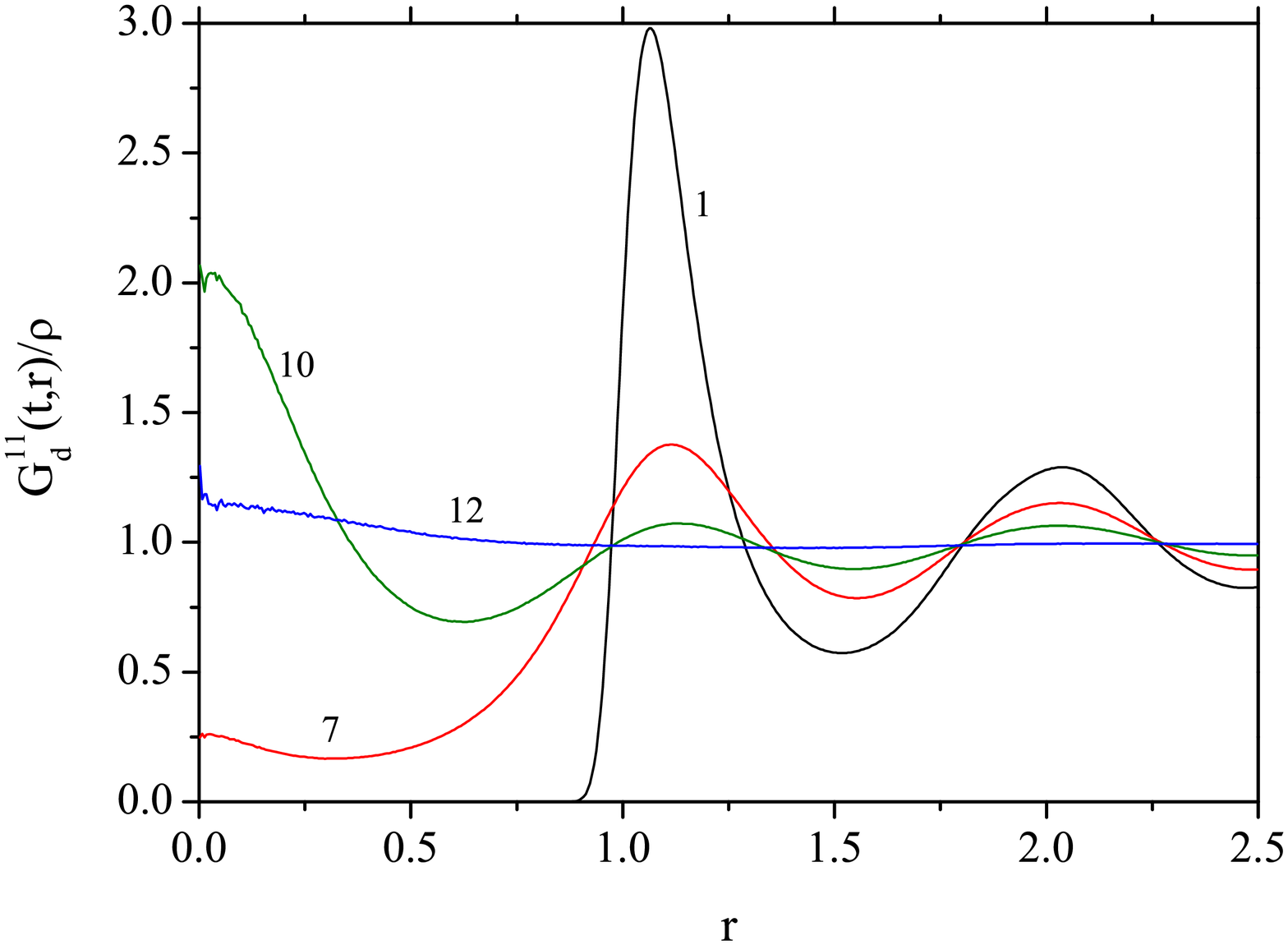}} 
\hfill\subfigure[]{\includegraphics[width=.49\columnwidth]{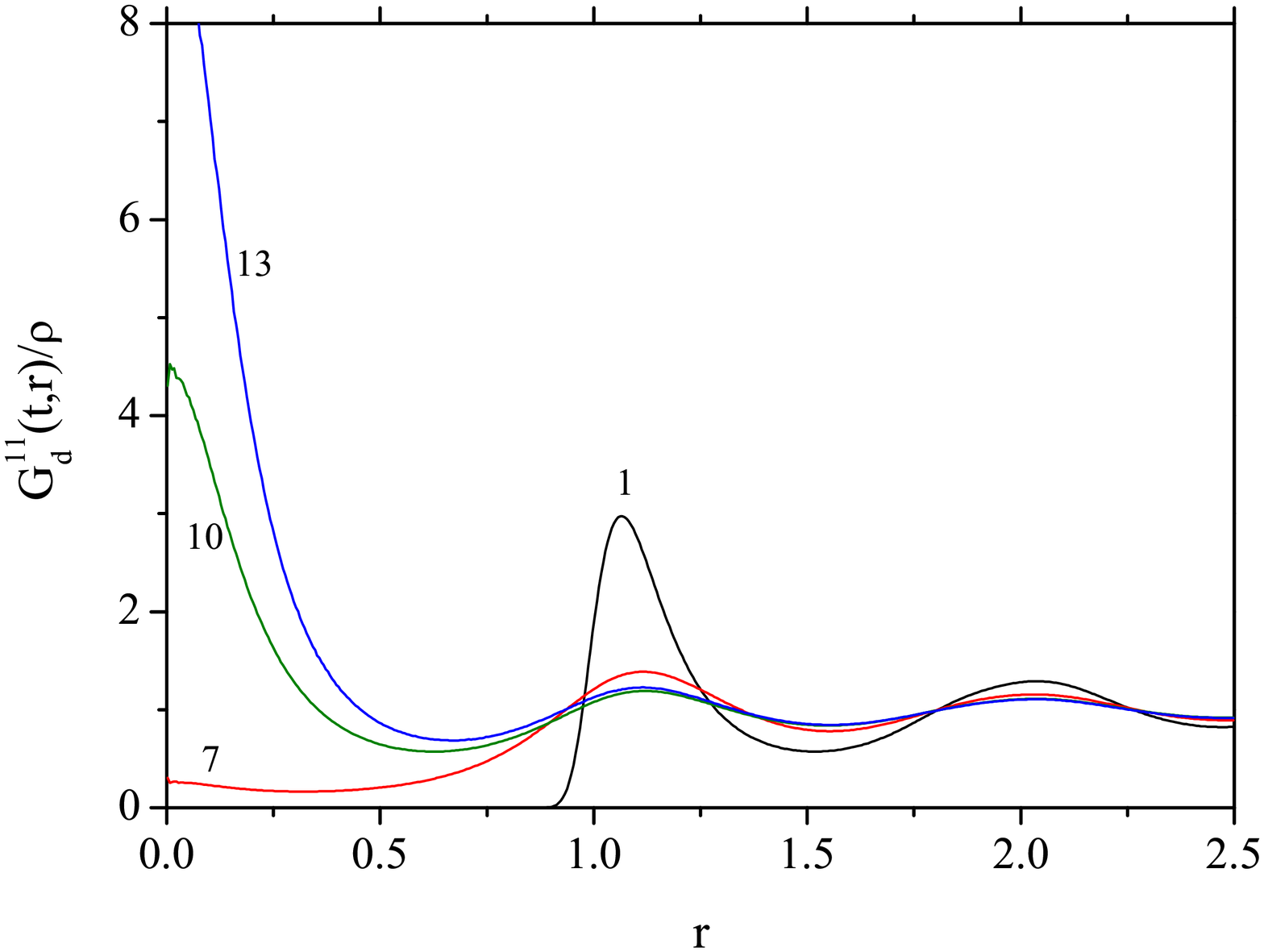}} 
\caption{ (Color online) Normalized distinct part of the VHCF, $G_d^{11}(r,t)/\rho$, as a function of $r$ at constant time $t$, with $x=0.2$ and $\mu=10^4$ (a) and $\mu=\infty$ (b).
The times are $t=2^i \times 0.015$, with $i$ given by the number next to each curve.
The density is $\rho=0.9$ and temperature $T=1$ in both cases.
\label{fig_GdL}}%
\end{figure}

\subsection{Shear viscosity \label{sec_visc}}

The total stress tensor of a one- or multicomponent system is given by \cite{allen}
\begin{equation}
\mathbf{\sigma}_{xy}=\sum_{i=1}^{N}\left[  m_{i}v_{i}%
^{x}v_{i}^{y}-\sum_{j>i}^{N}\frac{r_{ij}^{x}r_{ij}^{y}}{r_{ij}}\phi^{\prime
}\left(  r_{ij}\right)  \right]  .
\end{equation}
Now, the stress tensor of component $\alpha$ in a mixture can be defined as%
\begin{equation}
\mathbf{\sigma}_{xy}^{\alpha}=\sum_{i=1}^{N_{\alpha}}\left[  m_{i}v_{i}%
^{x}v_{i}^{y}-\sum_{j>i}^{N}\frac{r_{ij}^{x}r_{ij}^{y}}{r_{ij}}\phi^{\prime
}\left(  r_{ij}\right)  \right]  ,
\end{equation}
such that the total stress tensor is the sum of those of the two species,%
\begin{equation}
\mathbf{\sigma}_{xy}=\mathbf{\sigma}_{xy}^{1}+\mathbf{\sigma}_{xy}^{2}.
\end{equation}
Then one can write the stress-stress auto- and cross-correlation functions
$\eta_{\alpha\beta}\left(  t\right)  $ with $\alpha,\beta\in\left\{
1,2\right\}  $ as%
\begin{equation}
    \eta_{\alpha\beta}\left(  t\right)  =\left\langle \mathbf{\sigma}_{xy}%
    ^{\alpha}\left(  t\right)  \mathbf{\sigma}_{xy}^{\beta}\left(  0\right)
    \right\rangle , \label{etaacf}
\end{equation}
and thus the total correlation function is given by%
\begin{equation}
\eta\left(  t\right)  =\left\langle \mathbf{\sigma}_{xy}\left(  t\right)
\mathbf{\sigma}_{xy}\left(  0\right)  \right\rangle =\eta_{11}\left(
t\right)  +\eta_{22}\left(  t\right)  +2\eta_{12}\left(  t\right)  .
\end{equation}
The corresponding shear viscosities are computed via the Green-Kubo formula%
\begin{equation}
\eta_{\alpha\beta}=\frac{1}{Vk_{B}T}\int_{0}^{\infty}\eta_{\alpha\beta}\left(
t\right)  dt, \label{etaGK}%
\end{equation}
and the total shear viscosity of the mixture $\eta_m$ can be obtained as%
\begin{equation}
\eta_m=\eta_{1}+\eta_{2}+2\eta_{12}, \label{etam}
\end{equation}
where we write $\eta_{1}$ instead of $\eta_{11}$ and $\eta_{2}$ instead of
$\eta_{22}.$ The reason for separating $\eta_m$ into two contributions is that by doing so we can compare the results for finite $\mu$ with those for $\mu=\infty$. In this limit, the total viscosity goes to infinity while $\eta_1$ stays finite and approaches $\eta_1(\mu=\infty)$, the viscosity of the light particles moving through the porous matrix.

The numerical procedure we used to calculate Eq. $\left(\ref{etaGK}\right)$  was to store the values of
$\mathbf{\sigma}_{xy}^{\alpha}\left(  t\right)  ,$ $\mathbf{\sigma}%
_{yz}^{\alpha}\left(  t\right)  $ and $\mathbf{\sigma}_{xz}^{\alpha}\left(
t\right)$ at every third time step on disk, and perform the integration via a Fast Fourier
Transformation after the simulation, averaging over the three tensor elements in order to decrease statistical errors. In this way, it was not necessary to know the maximum integration time beforehand.

\begin{figure}[tb]
\subfigure[]{\includegraphics[width=.49\columnwidth]{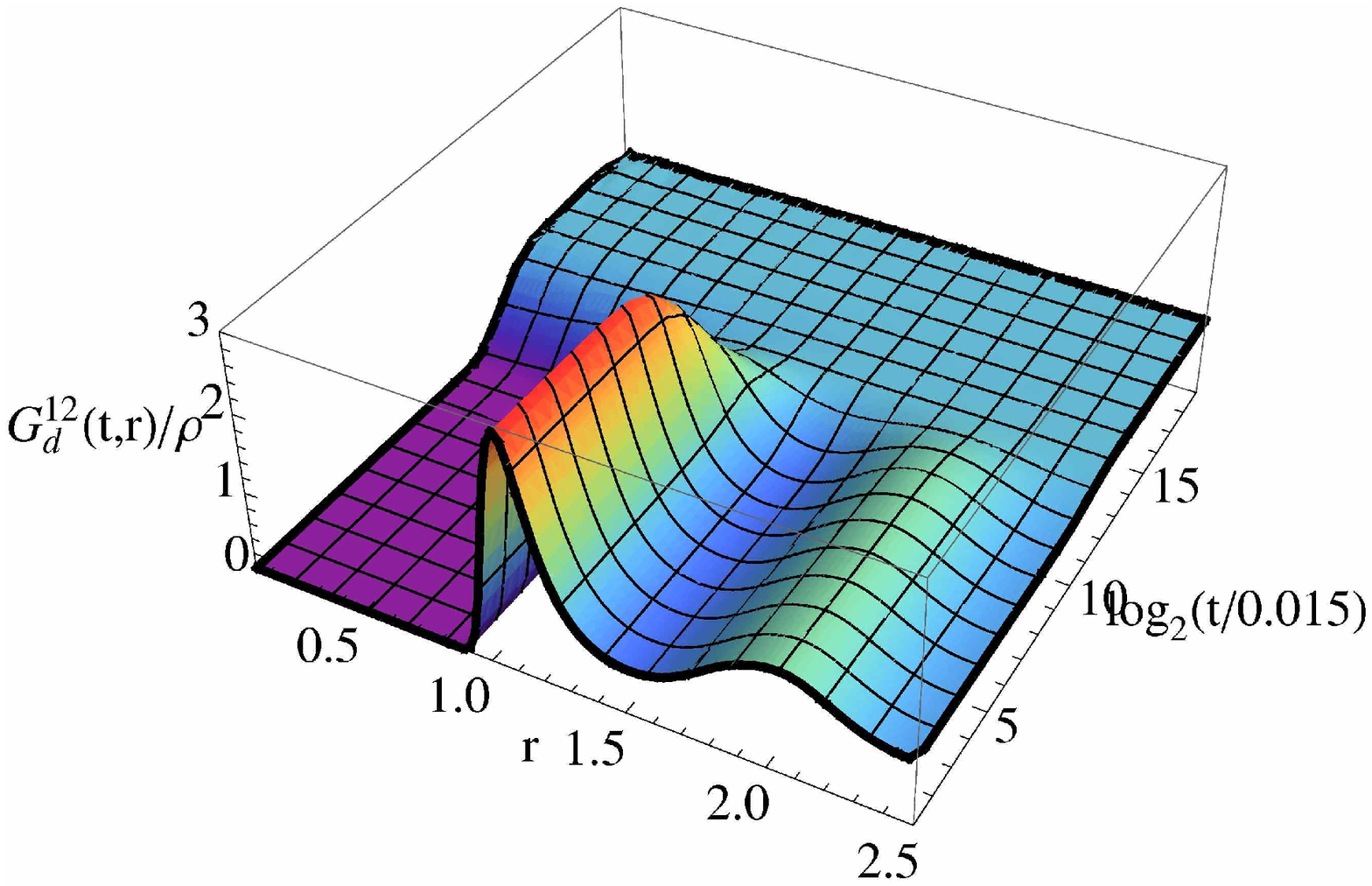}} 
\subfigure[]{\includegraphics[width=.49\columnwidth]{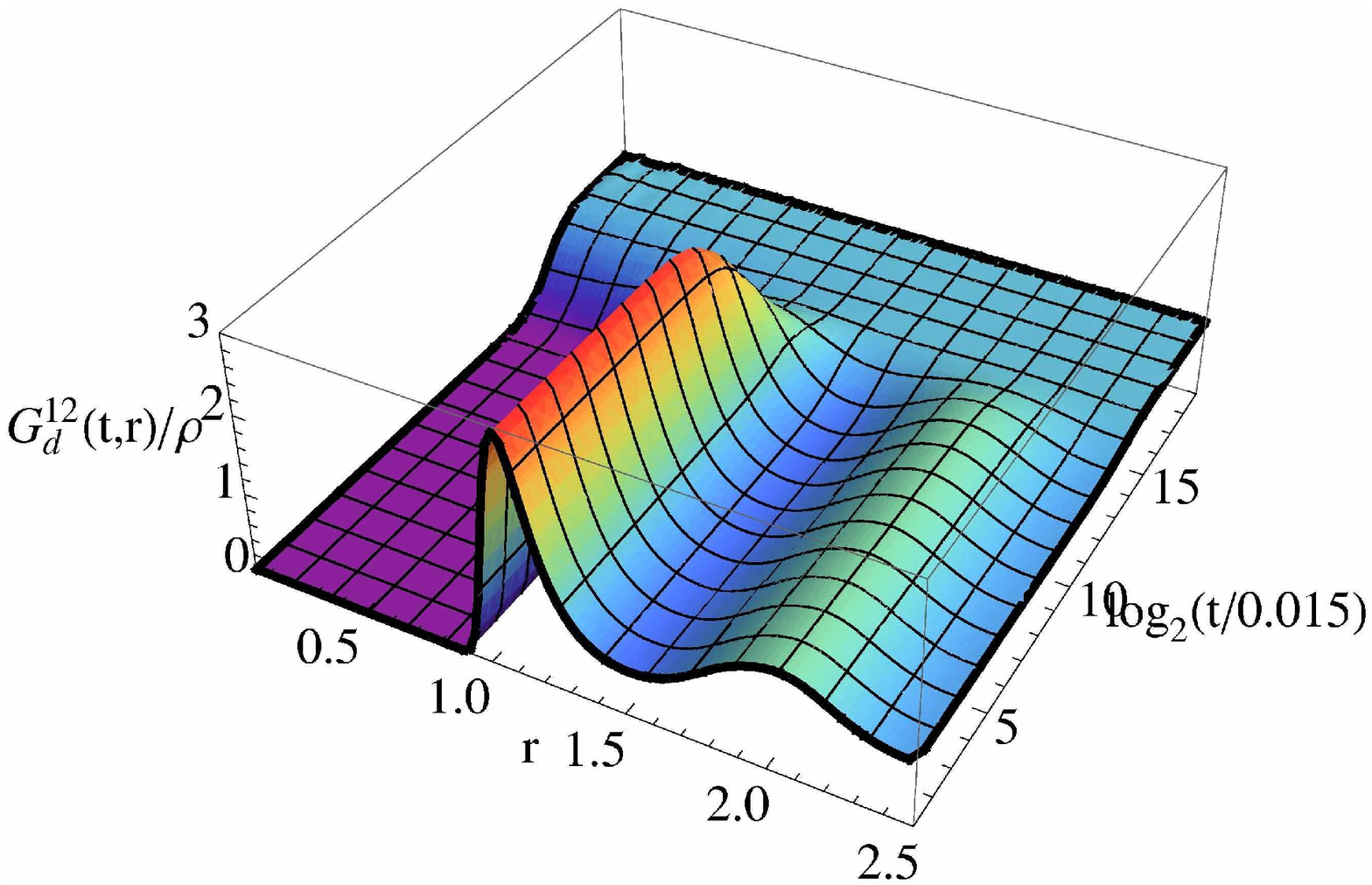}} 
\subfigure[]{\includegraphics[width=.5\columnwidth]{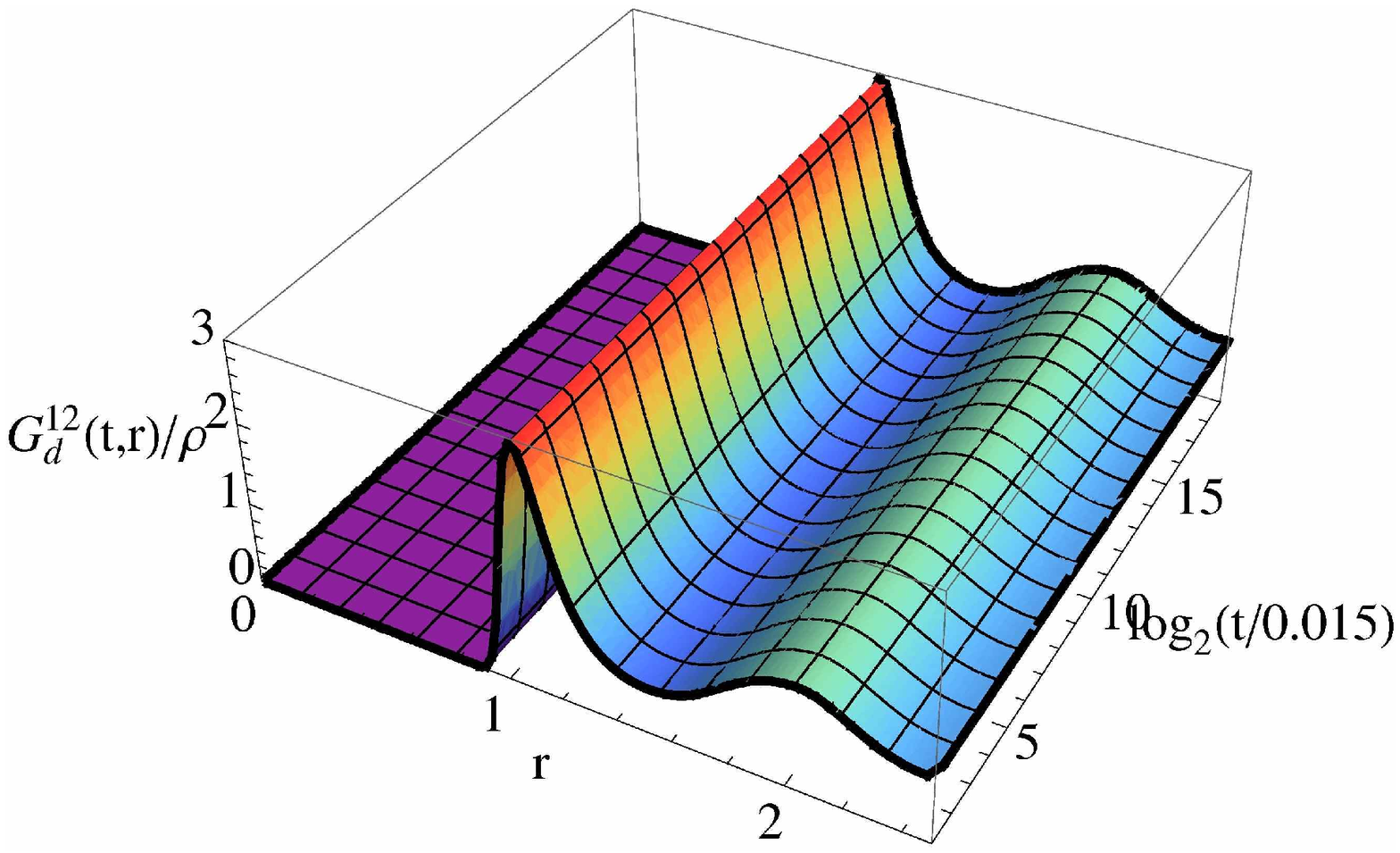}} 
\caption{ (Color online) Normalized distinct part of the van Hove function $G_d^{12}(r,t)/\rho$. The concentration of the heavy particles is $x=0.2$, and the mass ratio is $\mu=10^4$ (a), $\mu=3\times10^5$ (b) and $\mu=\infty$ (c). The density is $\rho=0.9$ and temperature $T=1$ in all cases.
\label{fig_Gd12}}%
\end{figure}

\subsection{Stokes-Einstein relation}

If we consider a macroscopic sphere ('tracer') of radius $R$ moving at a
velocity $\mathbf{V}$ in a liquid with shear viscosity $\eta$, according to
Stokes' law, the frictional force acting on the sphere is given by%
\begin{equation}
\mathbf{F}=-\zeta\mathbf{V,}%
\end{equation}
where $\zeta$ denotes the friction coefficient,
\begin{equation}
\zeta=C\pi\eta R,
\end{equation}
and the constant $C$ depends on the boundary conditions. In the case that
the viscous fluid sticks perfectly to the surface of the sphere (rough surface; \emph{stick} boundary condition), i. e. the fluid velocity is $\mathbf{v}=\mathbf{V}$
everywhere on the surface, $C$ is equal to $6.$ If, on the other hand, one
assumes that the fluid slips perfectly over the sphere (smooth surface; \emph{slip} boundary condition), the value obtained for $C$ is 4. In this case, only the normal component of the fluid velocity at the surface of the sphere is equal to that of the sphere velocity ($v_{\perp}=V_{\perp}$; no fluid can enter or leave the
sphere), and the tangential force at the surface is zero. These two values of $C$
can be found through purely hydrodynamic calculations, see for example \cite{LaLi,ZwaBix}.

\begin{figure}[tb]
\subfigure[]{\includegraphics[width=.49\columnwidth]{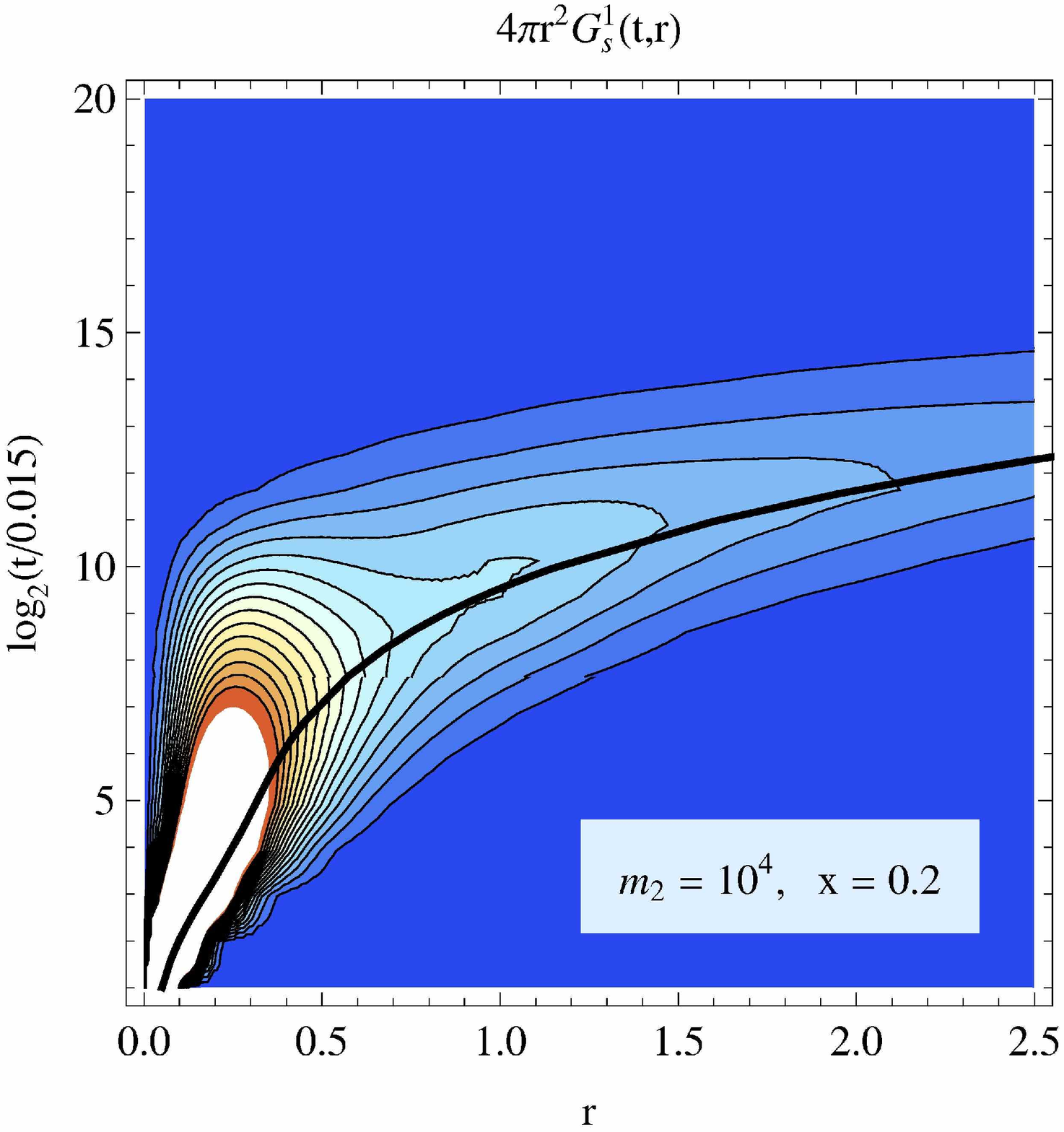}}
\hfill
\subfigure[]{\includegraphics[width=.49\columnwidth]{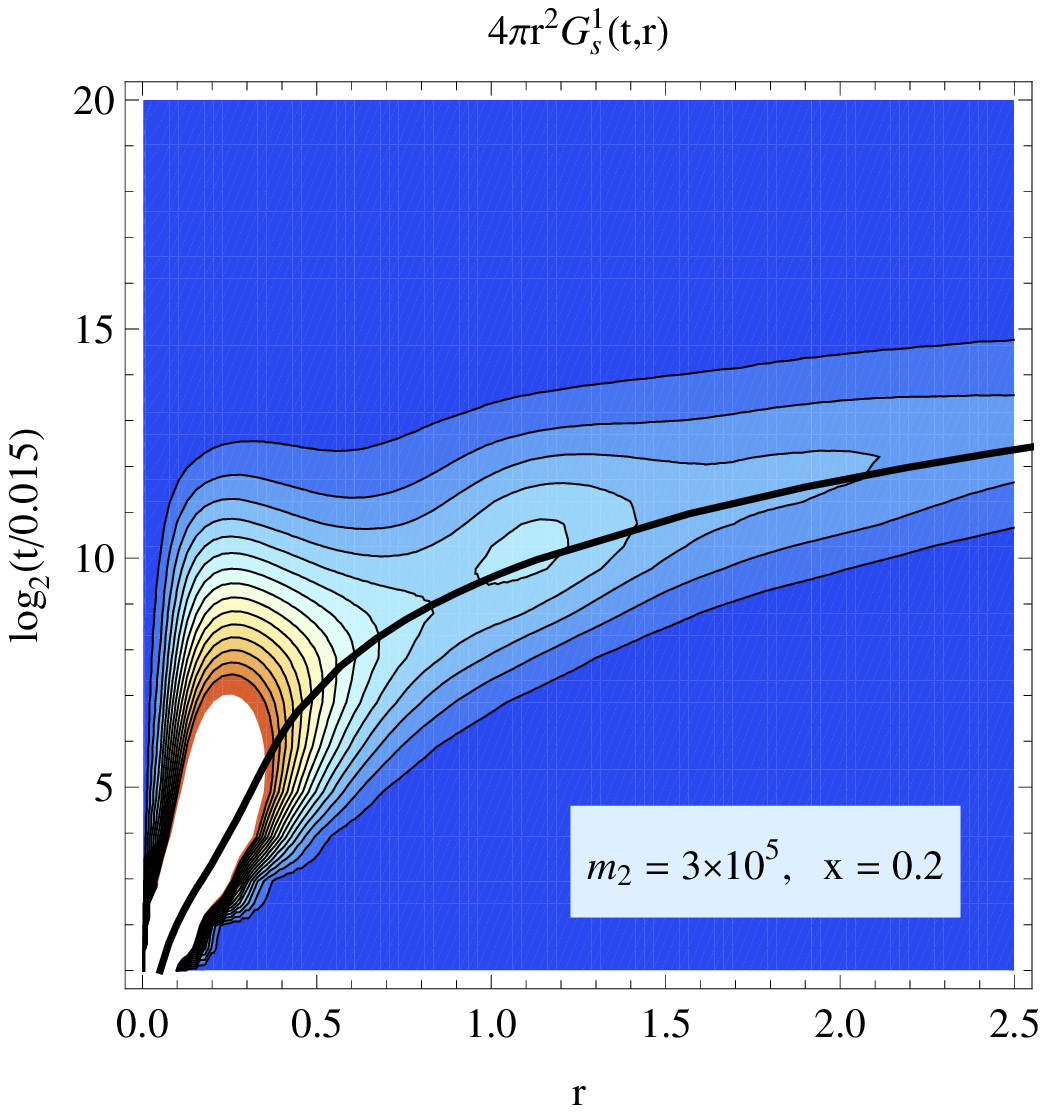}}

\subfigure[]{\includegraphics[width=.49\columnwidth]{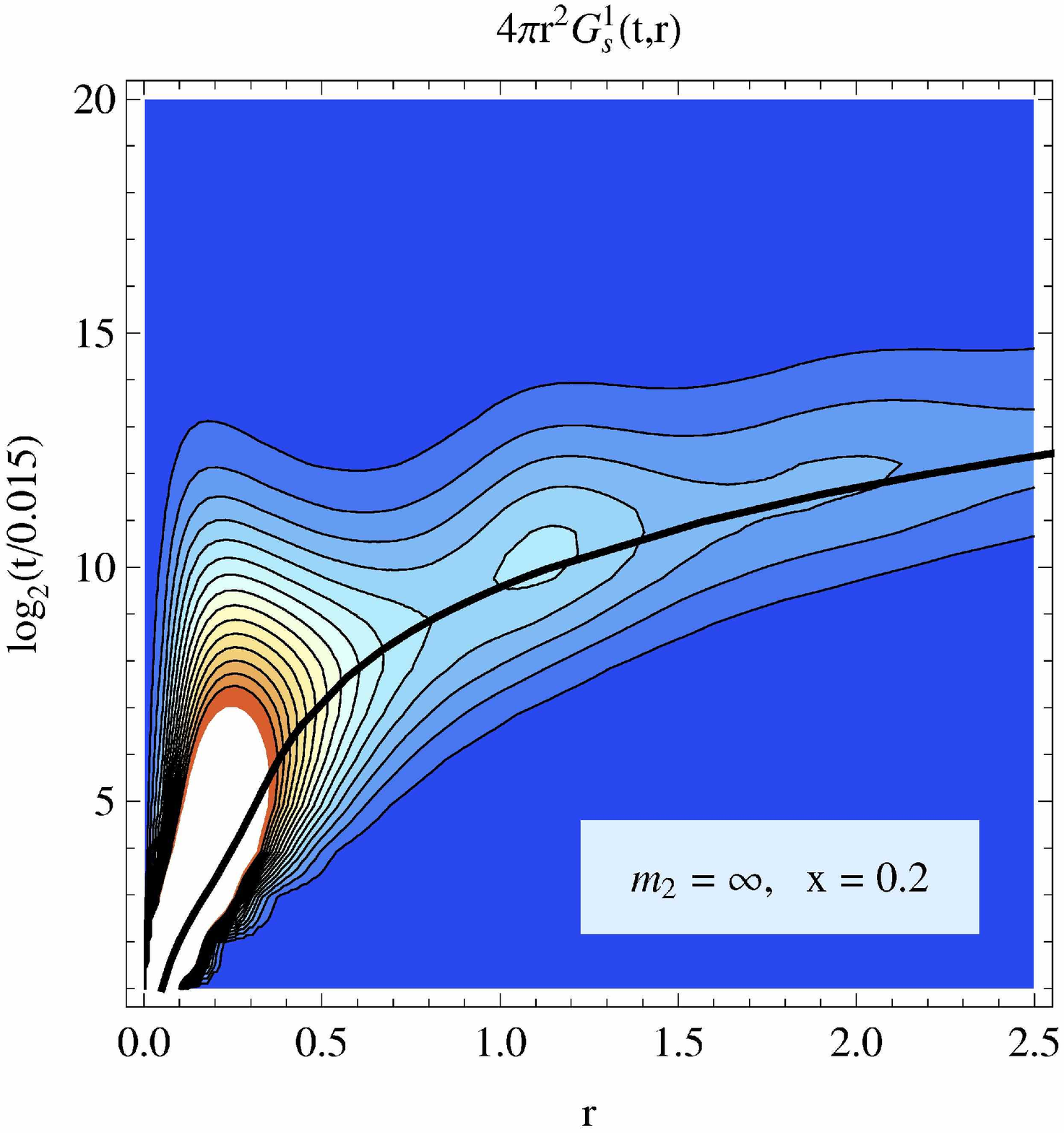}}
\hfill
\subfigure[]{\includegraphics[width=.49\columnwidth]{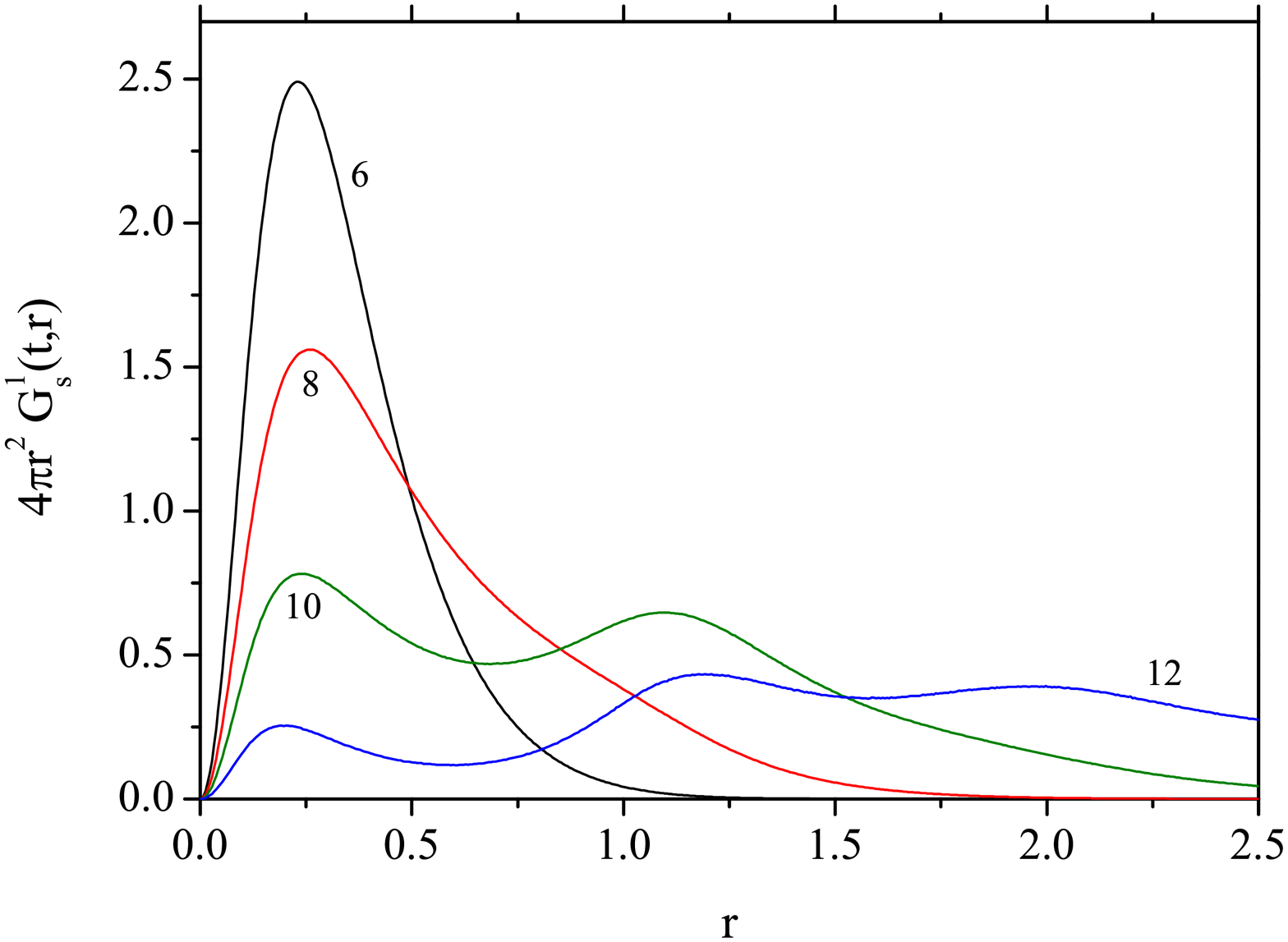}}
\caption{ (Color online) Self-part of the van Hove function for the light particles $G_s^{1}(r,t)$ times $4\pi r^2$. The concentration of the heavy particles is $x=0.2$, and the mass ratio is $\mu=10^4$ (a), $\mu=3\times10^5$ (b) and $\mu=\infty$ [(c),(d)]. The density is $\rho=0.9$ and temperature $T=1$ in all cases. Low values are shaded in blue, high values in red. The plot range in $z$-direction is [0,2], the difference between two contour lines is 0.12. Clipped regions are shown in white. The thick black lines correspond to the root mean square displacement $\sqrt{\left\langle \Delta r_{1}^2\left(  t\right)  \right\rangle}$.
The curves shown in (d) correspond to the times $t=2^i \times 0.015$, with $i$ given by the number next to each curve.
\label{fig_Gs1}}%
\end{figure}

Now, according to Einstein \cite{Einstein}, the friction coefficient $\zeta$ is
inversely proportional to the diffusion coefficient $D$ of the sphere with the thermal energy $k_B T$ as the constant of proportionality,
\begin{equation}
D=\frac{k_{B}T}{\zeta}. \label{SER}%
\end{equation}
Equation (\ref{SER}) is known as the Stokes-Einstein (SE) relation.
In the Brownian limit (tracer limit) of a binary isotopic mixture (only one particle of the heavy component), the Stokes-Einstein relation was also verified to hold in the form \cite{ould2,Schmidt}
\begin{equation}
D_{B}=\frac{k_B T}{C\pi\eta_{S}R_{H}}\label{SER2}
\end{equation}
where $D_B$ is the diffusion coefficient of the Brownian particle, $\eta_{S}$ is the shear viscosity of the solvent (light component), and $R_{H}$ is the so-called hydrodynamic radius.

\section{Results}

\subsection{Mean square displacement and cage effect}

\begin{figure}[tb]
\subfigure[]{\includegraphics[width=.9\columnwidth]{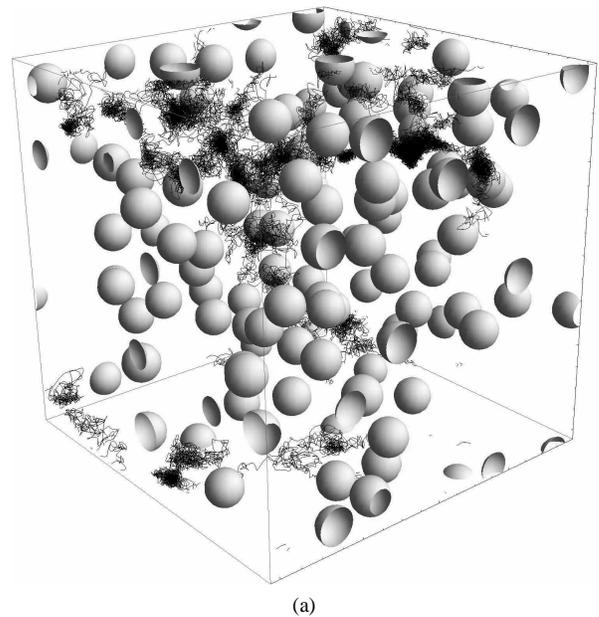}}
\hfill\subfigure[]{\includegraphics[width=.9\columnwidth]{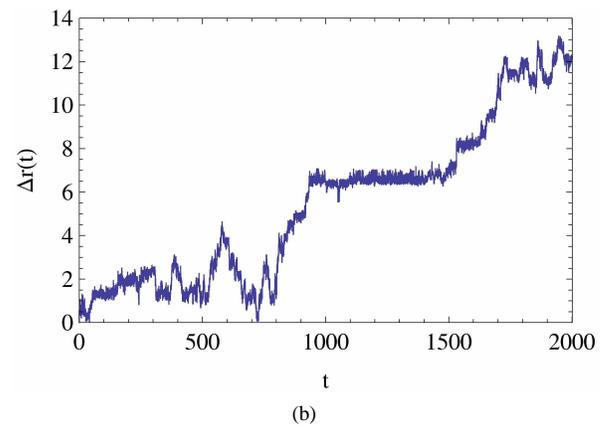}}
\caption{ (a) Trajectory of a particle of the light species in a system with fixed heavy particles ($\mu=\infty$), mole fraction $x=0.2$, density $\rho=0.9$ and temperature $T=1$. (b) displacement $\Delta r(t)$ of the same particle.
\label{fig_tra}}%
\end{figure}

We first want to focus our attention to the diffusion of the light particles.
Figure \ref{fig_MSD} shows the MSD of species 1 divided by $t$ in mixtures with $\mu=2$, $\mu=500$ and
$\mu=\infty$ (circles) for a density of $\rho=0.9$, concentration $x=0.2$ and temperature $T=1$, along
with the fitted MF curves [Eq. (\ref{M2}); solid lines]. The dashed horizontal lines indicate the values
of the diffusion coefficients times 6, obtained from independent calculations via the Green-Kubo formula
(\ref{DGK}). Obviously, there is consistency between the two calculation routes in the cases of finite
$\mu$, whereas for infinite mass ratio the MSD does not reach a linear-time behavior within the length
of the simulation. In general, we can distinguish three regimes of the MSD: the quadratic regime for
small times where the particles move ballistically at constant velocity, the linear regime with the
usual diffusion as given in Eq. (\ref{DEH}) for large times, and an intermediate region of anomalous
diffusion. Such a diffusion is often explained by the so-called 'cage effect' 
denoting the fact that particles are trapped inside a cage formed by their surrounding neighbors for
some time, before they can escape and diffuse in the usual way.
As seen from Fig. 1 the region of anomalous diffusion increases with $\mu$. This supports the idea \cite{schweizer}
that the trajectories of the light particles for large enough $\mu$ change from  relatively
smooth ones (Gaussian-like process) to intermittent ones with a large amplitude of displacements (highly non-Fickian process
or activated hopping). This cage effect can be seen in several
other quantities apart from the MSD.

\begin{figure}[tb]
\includegraphics[width=.9\columnwidth]{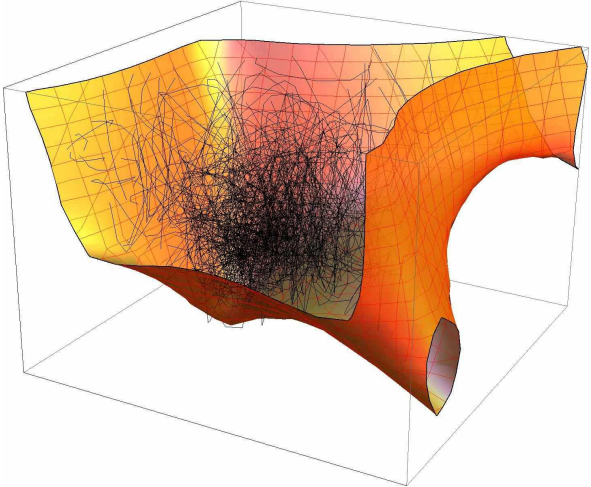}
\caption{Part of the trajectory shown in Fig. \ref{fig_tra} for $t$ between 1200 and 1350. The particle is trapped in a potential energy minimum. The isosurface corresponds to the total potential of all heavy particles. Inside the surface, the potential is lower than -5.
\label{fig_potmin}}%
\end{figure}

For example, the VACF of the light particles shows a distinct negative minimum, that becomes more pronounced if $\mu$ increases, as can be seen in Fig. \ref{fig_vacf} (a). In the VACF of the heavy particles, on the other hand, the minimum vanishes with increasing mass ratio [Fig. \ref{fig_vacf} (b)]. The position of the minimum $t_{\text{min}}$ offers a way to estimate the typical size of a cage. With a mean thermal velocity of $v_0=\sqrt{3}$ at $T=1$ the light particles on average travel a distance $v_0 t_{\text{min}}=0.26$ until they are reflected by the surrounding cage. Together with the particle radius 0.5 this results in a cage diameter of about 1.5.

Another indicator for the cage effect is the so-called non-Gaussian parameter (NGP) $\alpha_2$, considered as a measure of dynamic heterogeneity on intermediate time scales. Based on Eq. (\ref{MSDVHCF}) and the fact that the self-part of the VHCF takes a Gaussian form in the case of normal diffusion [Eq. (\ref{GaussVHCF})], one defines \cite{boon}
\begin{equation}
\alpha_{2}\left(  t\right)  =\frac{3\left\langle r^{4}\left(  t\right)
\right\rangle }{5\left\langle r^{2}\left(  t\right)  \right\rangle ^{2}}-1.
\end{equation}
Typically, $\alpha_2$ takes a value of $\sim0.1-0.2$ in the normal fluid regime \cite{Rahman}. Our simulation
results of the NGP for the same system as in Fig. \ref{fig_MSD} are shown in Fig. \ref{fig_alpha2}. As one can
see, with increasing mass the peak value of the NGP reaches almost 0.6, starting from the value 0.12 for $\mu=1$,
indicating that the cage effect becomes more pronounced for larger $\mu$. This is in accordance with the behavior
of the MSD seen in Fig. \ref{fig_MSD}.
It is also visible that for infinite mass $\alpha_2(t)$ is still nonzero at the largest times considered in our
simulations.

Finally, the cage effect should be observable via the VHCF. For normal diffusion, the value of the distinct VHCF
at the origin goes monotonically from 0 to $\rho$ with increasing time \cite{vanhove}. An additional peak
appearing at $r=0$ indicates that a particle has 'hopped' into the cage where the reference particle has
been at $t=0$, which by this time has escaped this cage formed by the surrounding particles \cite{kob}.
Figures \ref{fig_Gd11} - \ref{fig_Gd12} show the normalized distinct VHCFs $G_d^{11}(r,t)/\rho$ and
$G_d^{12}(r,t)/\rho$ for the same density, concentration and temperature as in Figs. \ref{fig_MSD}-\ref{fig_alpha2}
and large mass ratios $\mu=10^4, 3\times 10^5$ and $\infty$, where this effect is well seen.
At $\mu=10^4$ [Figs. \ref{fig_Gd11} (a) and \ref{fig_GdL} (a)], $G_d^{11}$ exhibits a peak at $r=0$ with about the
size of the first peak of the pair distribution function $g(r)$. With increasing mass ratio, both the height and the
width (in time-direction) of the peak grows, until for infinite mass ratio [Figs. \ref{fig_Gd11} (c) and
\ref{fig_GdL} (b)] the peak persists even up to the largest time, which is $0.015 \times 2^{19}$.
In $G_d^{12}$, shown in Fig. \ref{fig_Gd12}, the effect of increasing mass is apparent in the fact that the
structure of the pair distribution function $g(r)$ which is present at $t=0$ dissolves at later and later times,
until it remains unchanged throughout the whole simulation time for $\mu=\infty$ [Fig. \ref{fig_Gd12} (c)],
reflecting the 'frozen' configuration of the heavy particles.

But also the self-part of the van Hove function can reveal something about the cage effect and dynamic heterogeneity.
As we already mentioned, the usual shape for $G_s(r,t)$ is a Gaussian distribution in $r$ for any large enough time,
with its width increasing like $\sqrt{Dt}$.
This reflects a Fickian process. However, if the dynamics is strongly intermittent, it becomes heavily non-Gaussian
at intermediate time range.
For such an anomalous or hopping diffusion, the appearance of an additional peak is typical \cite{kob}. We show the
self part of the VHCF of the light particles $G_s^1(r,t)$, multiplied by $4\pi r^2$ to obtain the probability density,
in Fig. \ref{fig_Gs1}, for the same systems as in Fig. \ref{fig_Gd11} and \ref{fig_Gd12}. The formation of a
multi-peaked structure when $\mu$ takes large values ranging from $10^4$ to infinity is clearly visible in
Fig. \ref{fig_Gs1}. This Figure also demonstrates the three-peaked nature of $G_s^1$ for $\mu=\infty$.
The distance between the peaks indicates that the typical distance of two cages is about the size of a particle.

In order to examine the hopping behavior and cage entrapment more closely, we have taken a look at the
trajectory of a single light particle in the course of a simulation run with infinite mass ratio. The
mole fraction of the heavy species was again $x=0.2$, the density $\rho=0.9$, the temperature $T=1$, and the
number of particles used in the simulation was $N=500$. Figure \ref{fig_tra} (a) shows the obtained three-dimensional
path through the simulation box (the fixed heavy particles are depicted by gray spheres; periodic boundary conditions
apply), whereas Fig. \ref{fig_tra} (b) shows the distance $\Delta r(t)$ from the starting position at $t=0$ covered by
the tagged particle.
Both figures demonstrate that the particle is repeatedly trapped at some place, oscillating around a position with an
amplitude smaller than the particle size, before it hops again to some other trap. Hence this trajectory demonstrates
well the intermittent process, discussed above.
In Fig. \ref{fig_tra} (a) these traps appear as black regions where the trajectory passes many times.
In order to verify that the traps are indeed minima of the potential energy landscape created by the
fixed particles, we have plotted in Fig. \ref{fig_potmin} a magnified portion of the trajectory,
corresponding to the time period $1200 \lesssim t \lesssim 1350$, during which the particle is
trapped according to Fig. \ref{fig_tra} (b). It is obvious that for $\mu=1$ $\Delta r(t)$ in Fig.
8 (b) would show the known behavior of a pure fluid where
plateau ranges are practically absent. Only for large enough values
of the mass ratio and intermediate concentrations time intervals of
constant $\Delta r(t)$ become visible.
Also shown in the figure is a surface of constant
potential energy (the other mobile particles are not included in the calculation). It is apparent
that the surface forms a kind of bag, and the trajectory lies almost completely on the inside of it,
where the potential energy is smaller than on the outside. The dimensions of the portion are roughly
$\sigma \times \sigma \times \frac{\sigma}{2}$, and there is no heavy particle inside.

\begin{figure}[tb]
\includegraphics[width=.9\columnwidth]{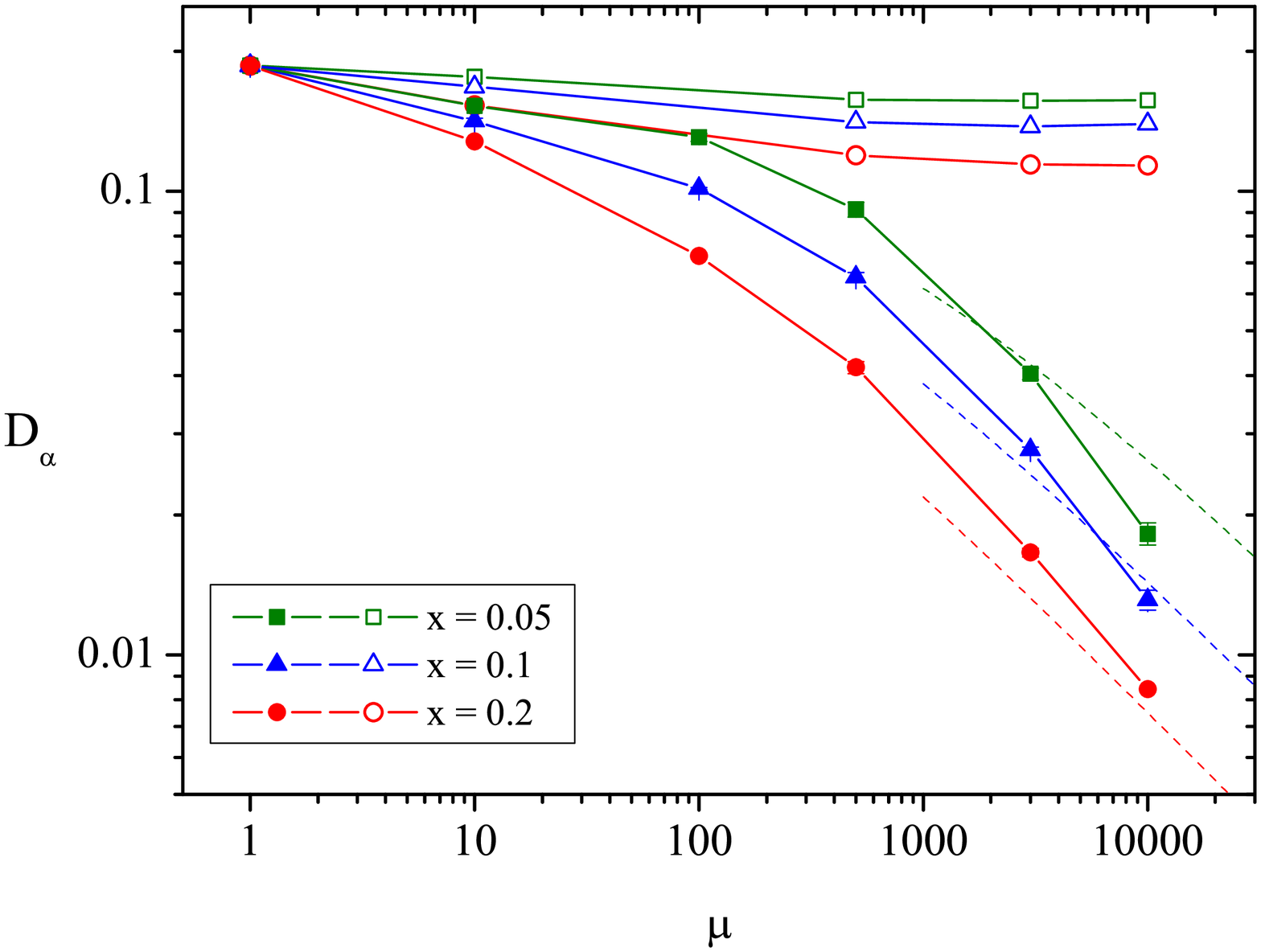}
\caption{ (Color online) Diffusion coefficient of the heavy (full symbols) and light species (open symbols)
as a function of the mass ratio $\mu$ for concentrations $x=0.05$ (green) $x=0.1$ (blue) and $x=0.2$ (red).
The dashed curves correspond to the linear model (see text). The density is $\rho=0.6$ and temperature $T=1.05$.
\label{fig_D12}}%
\end{figure}

\begin{figure}[tb]
\includegraphics[width=.9\columnwidth]{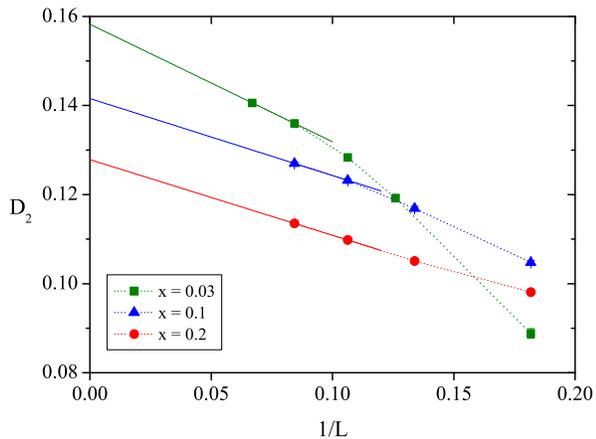}
\caption{ (Color online) System size dependence of the diffusion coefficient of the heavy species for
$\mu=10$ and different concentrations. Solid curves are extrapolations to $1/L=0$, dashed curves are
only guides to the eye. The density is $\rho=0.6$ and temperature $T=1.05$.
\label{fig_D2L}
}%
\end{figure}

\subsection{Diffusion coefficients}

Simulation results for the diffusion coefficients of light and heavy particles, $D_1$ and $D_2$ as
functions of the mass ratio $\mu$ are presented in Fig. \ref{fig_D12} for three different concentrations
$x=0.05$, 0.1 and 0.2. The temperature is $T=1.05$ and the density $\rho=0.6$. The results were corrected
for finite-size effects according to the equation \cite{fushiki,duenweg,yeh}
\begin{equation}
D(L)=D(\infty)-\frac{\alpha}{L} \label{DofL},
\end{equation}
where $\alpha$ is a fitting parameter.
Figure \ref{fig_D2L} shows the dependence of $D_2$ on the system size $L$ for some concentrations $x=0.2$, 0.1
and 0.03. In each case the mass ratio is $\mu=10$.
As one can see, the curves exhibit an increasing curvature with decreasing $x$, reflecting a departure
from the $1/L$-scaling behavior predicted by Eq. (\ref{DofL}).
Extrapolations to $1/L=0$ were performed using the two data points with the highest particle number in
each case.

Another difficulty that is caused by a small number of particles with high
mass is that the mean kinetic energies, and thus the temperatures, of the light
and heavy subsystems may deviate from each other. For a single Brownian
particle of mass $m_{B}$, this problem was studied in detail by Nuevo et al.
\cite{nuevo}. For a total number of particles $N$ and a mass
$m$ of the solvent particles, the mean square momentum of the Brownian
particle $\left\langle p_{B}^{2}\right\rangle $ will differ from its value in
the thermodynamic limit, $3Tm_{B}$, by a factor of%
\begin{equation}
f_B=\frac{N-1}{N-1+m_{B}/m}.\label{mfB}%
\end{equation}
Equation $\left(  \ref{mfB}\right),$ derived in \cite{espanol}, can be generalized to an arbitrary
number $N_{2}$ of heavy solute particles and $N_{1}=N-N_{2}$ light solvent particles, yielding%
\begin{equation}
f_{x}=\frac{N_{1}+\left(  N_{2}-1\right)  \mu}{N_{1}+N_{2}\mu}=\frac
{1-x+\left(  x-1/N\right)  \mu}{1-x+x\mu}.\label{mfx}%
\end{equation}
Table \ref{tab_Ekin} gives some examples of measured kinetic energies of the two species, $E_{1}^{kin}$
and $E_{2}^{kin}$, compared to the value predicted by Eq. $\left(\ref{mfx}%
\right)$. The thermodynamic limit value for this temperature is
$E_{kin}=\frac{3}{2}T=1.575.$ In some cases with low concentration and small system size the
deviations are found to reach up to 10\%.

\begin{table}[tb]
\begin{tabular}
[c]{lllllll}\hline\hline
$x$ & $\mu$ & $N$ & $E^{kin}_1$ & $E_2^{kin}$ & $\frac{E_2^{kin}}{1.575}$ & $f_x$ \\\hline\hline
0.2&    100 &250&   1.5801(4)   &1.5541(14)&    0.987   &0.981  \\
0.2&    100 &500&   1.5776(4)   &1.5646(15)&    0.993   &0.990\\
0.2&    100 &1000&1.5762(1) &1.5703(5) &    0.997 &0.995    \\

0.2&    3000&   250 &1.5748(7)&1.5757(15)&  1.000&  0.980   \\
0.2&    3000&   500 &1.5740(11)&1.5792(42)& 1.003&  0.990   \\
0.2&    3000&   1000&1.5764(2)&1.5695(9)&   0.996&  0.995   \\

0.02&   100 &250&   1.5784(1)   &1.4093(63)&    0.895&  0.866   \\
0.02&   100 &500&   1.5768(1)   &1.4893(40)&    0.946&  0.933   \\
0.02&   100 &1000   &1.5760(1)& 1.5291(16)& 0.971&  0.966   \\
0.02&   100 &2000   &1.5755(1)& 1.5515(16)& 0.985&  0.983   \\

0.02& 3000  &250&   1.5750(1)   &1.5737(72) &0.999& 0.803   \\
0.02&   3000&   500&    1.5753(2) &1.560(11)  &0.991&   0.902   \\
0.02&   3000&   1000&   1.5754(1)   &1.555(7)   &0.987& 0.951   \\
\hline\hline
\end{tabular}%
\caption{MD results for the kinetic energies $E^{kin}_1$ and $E^{kin}_2$ of the light and heavy subsystems,
for various concentrations, mass ratios and system sizes. The temperature is $T=1.05$ and the density $\rho=0.6$
in all cases.\label{tab_Ekin}}
\end{table}


\subsection{Shear viscosity}

\begin{figure}[tb]
\includegraphics[width=.9\columnwidth]{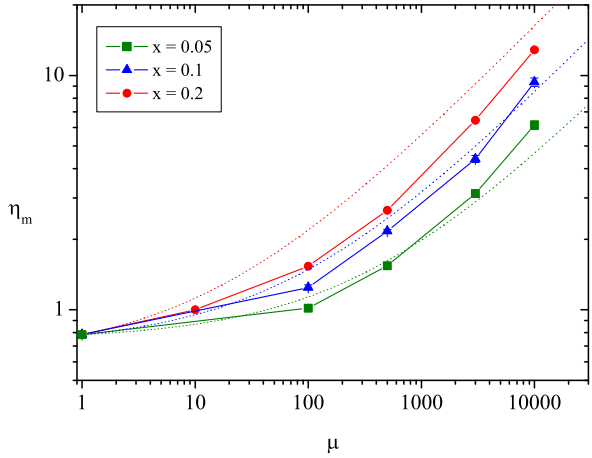}
\caption{ (Color online) Total shear viscosity $\eta_m$ as a function of the mass ratio $\mu$ for systems with $\rho=0.6$, $T=1.05$, and different mole fractions $x=0.2$, 0.1 and 0.05.
The dotted curves correspond to the linear model (see text).
\label{fig_eta}
}%
\end{figure}

\begin{figure}[tb]
\includegraphics[width=.9\columnwidth]{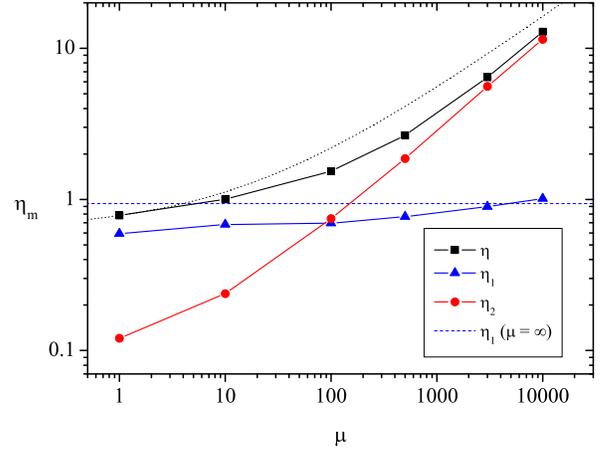}
\caption{ (Color online) Total shear viscosity $\eta_m$ and the contributions $\eta_1$ and $\eta_2$ of the two components as functions of the mass ratio $\mu$ for a system with $\rho=0.6$, $T=1.05$, and mole fractions $x=0.2$. The black dotted curve correspond to the linear model (see text), the blue dashed line indicates the value of $\eta_1$ obtained for $\mu=\infty$.
\label{fig_eta12}
}%
\end{figure}

For the same systems as in Fig. \ref{fig_D12} we have calculated the shear viscosities of the mixtures via
Eqs. (\ref{etaGK}) and (\ref{etam}). The results are presented in Figs. \ref{fig_eta} and \ref{fig_eta12}.
We compare them with a simple linear model assuming the mixture is ideal, and therefore the total viscosity
$\eta_{m}^{id}$ is given by
\begin{equation}
\eta_{m}^{id}=\left(  1-x\right)  \eta_{1}^{0}+x\eta_{2}^{0},
\end{equation}
where $\eta_{1}^{0}$ and $\eta_{2}^{0}$ denote the shear viscosities of the two components in their pure form.
Since the viscosity of a pure fluid scales with the square root of the mass of its particles, $\eta _{2}^{0}=\sqrt{\mu }\eta _{1}^{0}$, we have
\begin{equation}
    \frac{\eta _{m}^{id}}{\eta _{1}^{0}}=1+x\left( \sqrt{\mu }-1\right). \label{etamid}
\end{equation}
The dotted lines in Figs. \ref{fig_eta} and \ref{fig_eta12} were obtained from Eq. \ref{etamid}. Agreement with the MD data is in general quite good, only for $x=0.2$ the model overestimates the real values by up to 20\%. Figure \ref{fig_eta12} shows additionally the contributions $\eta_1$ and $\eta_2$ of the two mixture components as defined in section \ref{sec_visc}. It can be seen that while $\eta_m$ and $\eta_2$ are both increasing as $\sqrt{\mu}$ for large mass ratios, $\eta_1$ is growing only slowly and reaches the value obtained for $\mu=\infty$ (blue dashed line) at $\mu=10^4$.

\begin{figure}[tb]
\includegraphics[width=.9\columnwidth]{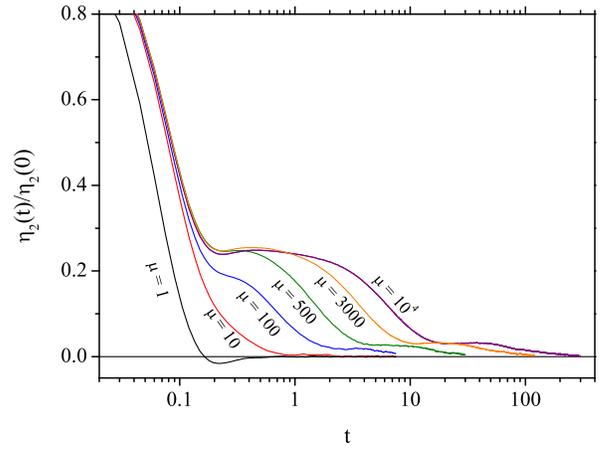}
\caption{ (Color online) Normalized stress-stress autocorrelation function $\eta_2(t)$ for the heavy component of the system with $x=0.2$ of Fig. \ref{fig_eta}.
\label{fig_etaacf}}%
\end{figure}

In Fig. \ref{fig_etaacf}, we plot several stress-stress autocorrelation functions $\eta_2(t)\equiv\eta_{22}(t)$ [see Eq. (\ref{etaacf})] for the systems with $x=0.2$ and various values of the mass ratio $\mu$. It is obvious that with increasing $\mu$ the relaxation times grow in a similar manner as we observed for the VACF $\psi_2(t)$. Consequently, the numerical integration of Eq. (\ref{etaGK}) has to be extended up to very large times $t_{\text{max}}\sim 100$ in order to reach the plateau value of $\eta_2$.

%

\begin{table}[tb]
\begin{tabular}
[c]{llllll}\hline\hline
$x$ & $\mu$ & $D_{2}$ & $\eta_{\text{m}}$ & $\eta_{\text{1}}$ &
$\eta_{\text{2}}$\\\hline
0.2 & 1 & 0.1865(7) & 0.786(12) & 0.593(9) & 0.121(3)\\
0.2 & 10 & 0.1278(5) & 1.001(9) & 0.683(7) & 0.238(5)\\
0.1 & 10 & 0.142(2) &  &  & \\
0.05 & 10 & 0.154(2) &  &  & \\
0.03 & 10 & 0.158(3) &  &  & \\
0.02 & 10 & 0.169(3) &  &  & \\
0.2 & 100 & 0.0725(5) & 1.536(18) & 0.696(12) & 0.746(13)\\
0.1 & 100 & 0.1013(6) & 1.246(31) & 0.758(22) & 0.428(14)\\
0.05 & 100 & 0.131(2) & 1.016(18) & 0.768(9) & 0.216(4)\\
0.02 & 100 & 0.151(3) &  &  & \\
0.2 & 500 & 0.042(1) & 2.660(67) & 0.769(16) & 1.864(61)\\
0.1 & 500 & 0.0652(15) & 2.165(40) & 0.798(32) & 1.269(38)\\
0.05 & 500 & 0.091(3) & 1.544(28) & 0.769(22) & 0.762(16)\\
0.02 & 500 &  & 1.088(26) &  & \\
0.2 & 3000 & 0.0166(3) & 6.44(22) & 0.895(28) & 5.61(19)\\
0.1 & 3000 & 0.0277(3) & 4.41(15) & 0.779(36) & 3.69(12)\\
0.05 & 3000 & 0.040(1) & 3.14(11) & 0.795(29) & 2.385(82)\\
0.02 & 3000 & 0.056(2) & 1.896(34) & 0.780(14) & 1.076(22)\\
0.2 & $10^{4}$ & 0.00844(3) & 12.87(40) & 1.013(60) & 11.46(36)\\
0.1 & $10^{4}$ & 0.01313(7) & 9.37(39) & 0.78(4) & 8.31(35)\\
0.05 & $10^{4}$ & 0.0194(15) & 6.14(26) &  & \\
0.02 & $10^{4}$ &  & 3.85(17) &  & \\
0.01 & $10^{4}$ &  & 2.25(9) &  & \\\hline\hline
\end{tabular}%
\caption{MD results for the diffusion coefficients and viscosities, obtained via the Green-Kubo formulas (\ref{DGK}) and (\ref{etaGK}), for various concentrations and mass ratios. The temperature is $T=1.05$ and the density $\rho=0.6$ in all cases.\label{tab_Deta}}
\end{table}

\subsection{Stokes-Einstein relation}

\begin{figure}[tb]
\includegraphics[width=.9\columnwidth]{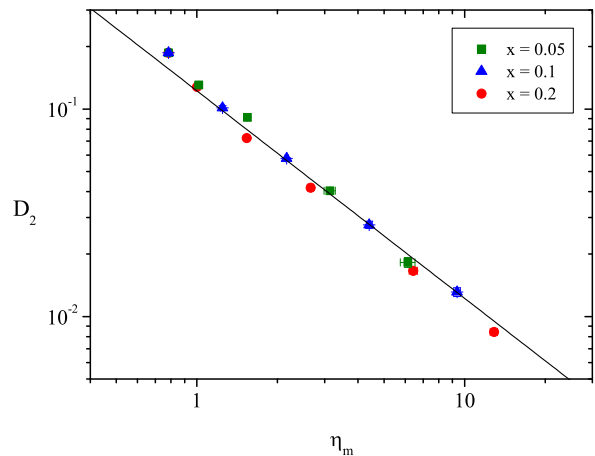}
\caption{ (Color online) Fit of the MD data to Eq. (\ref{SEfit}) for concentrations $x=0.2$, 0.1 and 0.05 and different mass ratios $\mu=1,10,100,500,3000,10^{4}$. The straight line corresponds to an exponent of -1, and the fit yields a hydrodynamic radius $R_{H}=0.68$ (assuming slip boundary conditions). The density is $\rho=0.6$ and temperature $T=1.05$.
\label{fig_SEfit}
}%
\end{figure}

In order to look for a relation between the diffusion coefficient of the heavy
particles and the shear viscosity, we plot $D_{2}$ from Fig. \ref{fig_D12} and
$\eta_{m}$ from Fig. \ref{fig_eta} in a double-logarithmic scale in Fig.
\ref{fig_SEfit} (full symbols). We observe that these data points lie close to
a straight line, which might be represented by the equation%
\begin{equation}
D_{2}=A\eta_{m}^{-\alpha}, \label{SEfit}%
\end{equation}
where $A$ and $\alpha$ are fitting parameters (a similar relation was also
suggested in \cite{ali}). Indeed, a linear fit leads to the values
$\alpha=1.07$ and $A=0.131.$ Thus, we propose a Stokes-Einstein-like relation
with $\alpha=1$ which yields a value of $A=0.122$ (solid line in Fig. \ref{fig_SEfit}). 
From such a relation one may extract an effective hydrodynamic radius $R_{H}$ by
identifying $A=k_{B}T/C\pi R_{H}$ according to Eq. $\left(\ref{SER2}\right).$ Assuming slip boundary conditions, we obtain $R_{H}=0.68,$ which seems to be
reasonable for our interaction potential. Similar values have also been found before, e. g. in \cite{McPhie}.

Equation (\ref{SEfit}) also allows us to apply the linear ideal-mixture model (\ref{etamid}) for $\eta_m$ to the diffusivity $D_2$. Combining the two equations yields
\begin{equation}
    D_{2}\left(x, \mu \right) =\frac{B}{1+x\left(\sqrt{\mu }-1\right) }, \label{D2id}
\end{equation}
with $B=A/\eta _{1}^{0}=0.156.$ The curves resulting from Eq. $\left( \ref{D2id}\right) $ for $x=0.05,$ 0.1 and 0.2 are shown by the dashed lines in Fig. \ref{fig_D12}.

\begin{figure}[tb]
\subfigure[]{\includegraphics[width=.9\columnwidth]{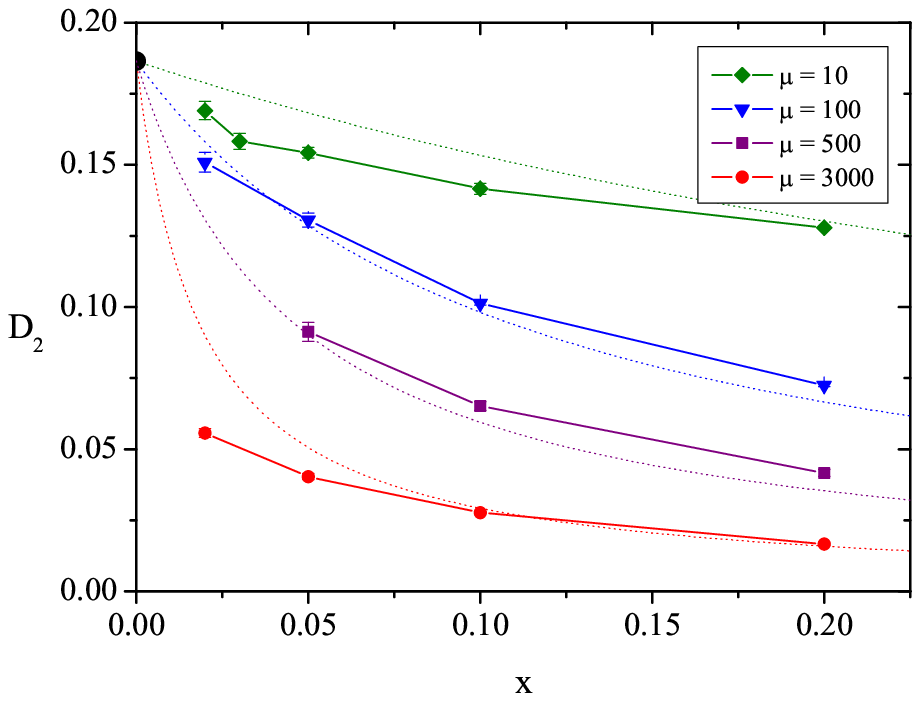}}

\subfigure[]{\includegraphics[width=.9\columnwidth]{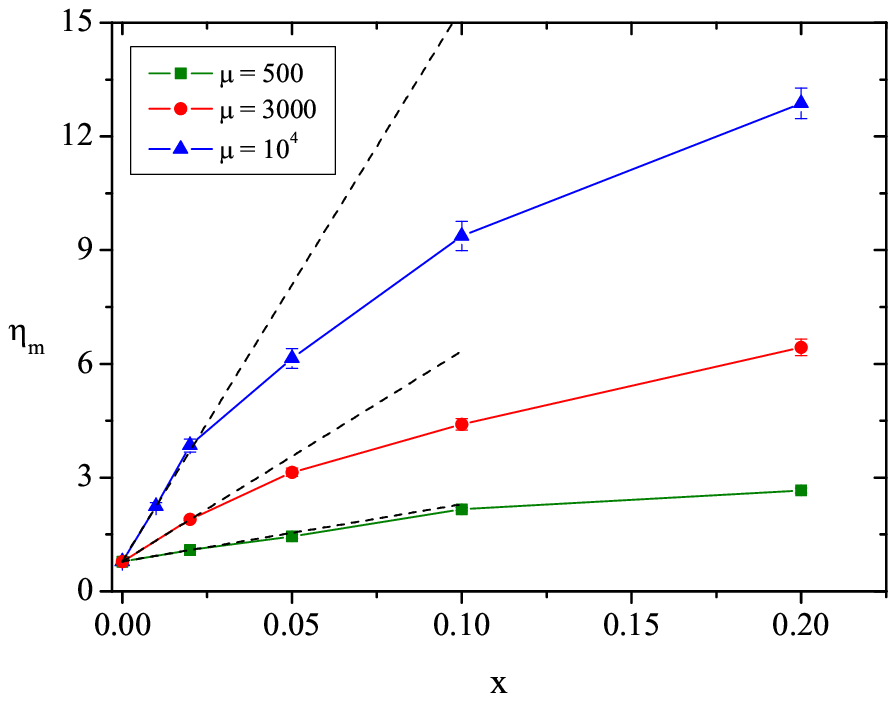}}
\caption{ (Color online) Dependence of the diffusion coefficient of the heavy species (a) and the shear viscosity of the mixture (b) on the concentration for different mass ratios.
The dotted curves in (a) correspond to the linear model (see text), the dashed curves in (b) are linear fits.
The density is $\rho=0.6$ and temperature $T=1.05$.
\label{fig_D2eta}}%
\end{figure}

\subsection{Concentration dependence}

Finally, we investigated the dependence of $D_2$ and $\eta_m$ on the concentration. Figure \ref{fig_D2eta} (a) shows the diffusivity of the heavy component as a function of $x$ for $\rho=0.6$, $T=1.05$, and $\mu=10,$ 100, 500 and 3000.
For comparison, we also include the curves predicted by the linear model, namely Eq. $\left( \ref{D2id}\right) $ at fixed values of $\mu$. In this case, however, we set $B=D_2(\mu=1)=0.187$, since otherwise the value of $D_2(x\rightarrow0)$ predicted by the SE-relation $(\ref{SEfit})$ is too low, which is also apparent from Fig. \ref{fig_SEfit}.
It seems that for not too small values of $x$, the linear model describes the behavior quite well, while for small concentrations the deviations are getting larger. Also, from the MD data it is not clear whether $D_2$ approaches the same value for any $\mu$ as $x$ goes to zero. Since finite-size effects increase when approaching the Brownian limit, we could not answer this question.

A clearer picture can be given regarding the concentration dependence of the shear viscosity, shown in Fig. \ref{fig_D2eta} (b). For $x\rightarrow0$, $\eta_m$ of course approaches the pure-fluid value $\eta_1^0$ of the light component, which is 0.786(12) for the chosen density and temperature. At small concentrations, the function $\eta_m(x,\mu)$ can be approximated by a linear ansatz,
\begin{equation}
\frac{\eta_{m}\left(  x,\mu\right)  }{\eta_{1}^{0}}=1+k_{\eta}\left(
\mu\right)  x+\ldots\label{etaxmu},
\end{equation}
with a $\mu$-dependent coefficient $k_{\eta}$. Comparison with Eq. (\ref{etamid}) yields
$k_{\eta}\left(\mu\right)=\sqrt{\mu-1}$ for the simple linear model. The slopes obtained from the simulation data ($k_{\eta}=19$ for $\mu=500,$ $k_{\eta}=70$ for $\mu=3000$ and
$k_{\eta}=186$ for $\mu=10000$) are in qualitative agreement with this assumption. In any case, the observed numbers are much larger than the well-known value of 2.5 proposed by Einstein for a suspension of solid particles in a liquid at small concentrations \cite{Einstein,LaLi2}.

\section{Conclusion}

We have performed extensive MD simulations of binary Lennard-Jones fluids whose components are identical except for their mass, such that only dynamic properties like transport coefficients and time correlation functions change with varying mass ratio $\mu$ and concentration $x$ of the two species. In particular, we have studied diffusion coefficient, shear viscosity, velocity and stress-stress autocorrelation functions, the van Hove space-time correlation function and the mean-square displacement for a range of (small) mole fractions of the heavy component, and high mass ratios up to infinity. The latter case was realized by fixing the heavy particles at their starting positions during the whole simulation run.

We found that especially at high liquid densities and high mass ratios the large difference in relaxation times of light and heavy particles leads to a pronounced cage effect for the light component. It can be observed as an intermediate region of anomalous diffusion in the mean-square displacement, a large maximum of the non-Gaussian parameter,
and additional peaks in both the self- and distinct part of the van Hove correlation function. When tracing the trajectory of a single light particle, it turns out that its motion is characterized by hopping between separate local minima of the potential energy landscape. Thus our study gives in fact an example that a rather stable 'solvent cage' can be formed in mixtures just because of a strong mass asymmetry effect.

Furthermore, we established a generalized Stokes-Einstein relation between the diffusion coefficient of the heavy component and the total shear viscosity of the mixture that is valid in the whole range of mass ratios and concentrations. In order to obtain accurate results, it was necessary to correct for the system size dependence of the diffusivity, and to ensure that the Green-Kubo integral for the shear viscosity has reached its plateau value.

Mass dependence of both viscosity and diffusivity are approximately predicted by a simple linear model assuming an ideal mixture behavior. For small concentrations, the shear viscosity follows a linear dependence on $x$ with a slope going roughly as $\sqrt{\mu}$, whereas for the diffusion coefficient of the heavy species due to computational limitations no conclusive result could be obtained.

There are several possible ways of extending the results presented here. Apart from studying other transport coefficients such as thermal conductivity or mutual diffusion, it would be interesting to perform a similar investigation close to the critical point of the phase diagram. Furthermore it is planned to study the percolation threshold at high concentrations of the heavy component, and to leave the hydrodynamic regime and look at the wave-vector dependence of the dynamic quantities (see, e.g., \cite{Bryk}). For the latter problem, calculations have already been performed the results of which will be published elsewhere.

\begin{acknowledgments}
We acknowledge support by the Fonds zur F\"orderung der
wissenschaftlichen Forschung under Project No. P18592-TPH. We also thank I. Omelyan for helpful discussions.
\end{acknowledgments}

\bibliography{fenz}
\end{document}